\documentclass[%
 reprint,
 amsmath,amssymb,
 aps,
floatfix,
]{revtex4-2}

\usepackage{ragged2e}

\usepackage{graphicx}
\usepackage{dcolumn}
\usepackage{bm}

\usepackage{setspace}
\usepackage{amsmath}
\usepackage{physics}
\usepackage{graphicx}
\usepackage{float}
\usepackage{caption}
\usepackage{subcaption}
\usepackage{siunitx}
\usepackage{threeparttable}
\usepackage{mathtools}
\usepackage{booktabs}
\usepackage{makecell}

\newcommand{\cavg}{\ensuremath{\langle c \rangle}}
\newcommand{\avg}[1]{\ensuremath{\langle #1 \rangle}}
\newcommand{\barep}{\ensuremath{\overline{\ve}}}
\newcommand{\bart}{\ensuremath{\overline{t}}}
\newcommand{\ve}{\ensuremath{\varepsilon}}

\begin{document}

\preprint{APS/123-QED}

\title{Physical limits on chemical sensing in bounded domains}

\author{Daniel R. McCusker}
 \affiliation{Applied Physics Graduate Program, University of Michigan}
\author{David K. Lubensky}%
 \email{dkluben@umich.edu}
\affiliation{%
 Department of Physics, University of Michigan
}%

\date{\today}

\begin{abstract}
Cells respond to chemical cues, and the precision with which they can sense these cues is fundamentally limited by the stochastic nature of diffusion and ligand binding. Berg and Purcell famously investigated how well a small sensor in an infinite ligand bath can determine the ligand concentration, and a number of subsequent analyses have refined and built upon their classical estimates.  Not all concentration sensing problems, however, occur in such an infinite geometry. At different scales, subcellular sensors and cells in tissues are both often confronted with signals whose diffusion is affected by confining boundaries.  It is thus valuable to understand how basic limits on chemosensation depend on the sensor's size and on its position in the domain in which ligand diffuses. Here we compute how sensor size and proximity to reflecting boundaries affect the diffusion-limited precision of chemosensation for various geometries in one and three dimensions. We derive analytical expressions for the sensing limit in these geometries. Among our conclusions is the surprising result that, in certain circumstances, smaller sensors can be more effective than larger sensors. This effect arises from a trade-off between spatial averaging and time averaging that we analyze in detail.  We also find that proximity to confining boundaries can degrade a sensor's precision significantly compared to the precision of the same sensor far from any boundaries.
\end{abstract}

\maketitle


\section{\label{sec:intro}Introduction}

Cells perceive and respond to chemical cues. The precision with which these cues can be sensed is affected by the random, diffusive motion of discrete ligand molecules and by stochastic interactions between ligands and receptors. Berg and Purcell argued in early, seminal work that a cell's ability to estimate a chemical concentration is fundamentally limited by shot noise in the diffusive arrival of ligands, independent of any detailed sensing mechanism. Considering idealized sensor models and physical heuristics, they argued that a sensor sitting in an infinite ligand bath can estimate a background ligand concentration $\langle c \rangle $ with a fractional variance no smaller than
\begin{align}
\label{eq:bp-limit}
    \frac{\delta c^2}{\langle c \rangle ^2} &\sim \frac{1}{Da \langle c \rangle  T}
\end{align}
where $a$ is the sensor's linear size, $D$ is the ligand diffusion constant, and $T$ is the time over which the sensor averages its measurements \cite{Berg1977}. 

Berg and Purcell calculated $\delta c^2/\avg{c}^2$ for two simplified models: the ``perfect instrument," in which the cell is modeled as a permeable volume which takes instantaneous snapshots of the number of molecules within it, and the perfect absorber, in which a cell detects, then absorbs and destroys, all diffusing molecules incident upon it and estimates the concentration of ligands from their arrival rate.  They showed that both of these cases obey the scaling of Eq. \ref{eq:bp-limit}, albeit with different numerical prefactors \cite{Berg1977, Endres2008}.

Subsequent developments have examined in more depth how these results change when explicit ligand-receptor binding is included. 
Bialek and Setayeshgar, for example, used the fluctuation-dissipation theorem (FDT) together with mean-field, mass action binding kinetics to find the variance in the fraction of ligand-bound receptors. They concluded that the sensing limit always has a term that scales like Eq. \ref{eq:bp-limit}, with $a$ representing the linear size of a receptor cluster; this term gives the unavoidable bound on sensing precision due to diffusion of the ligand. Noise in ligand-receptor interactions can then contribute additional additive terms to $\delta c^2/\avg{c}^2$ \cite{Bialek2005}.  This basic structure persists even when cooperative interactions among multiple receptors are present \cite{Bialek2008}. Moreover, for typical biological parameters, sensing systems operate in the diffusion-limited regime, with relatively smaller contributions from the terms that account noise in binding and unbinding \cite{RTW2016}.  Further work has addressed the effect of correlations induced by repeated binding and unbinding of the same ligand molecule, which appear to introduce additional factors of $1-\overline{p}$, where $\overline{p}$ is the occupancy probability of a single receptor \cite{Berezhkovskii2013,Kaizu2014}; investigators have also applied more sophisticated maximum likelihood methods to the problem of concentration estimation \cite{Endres2009}.

Importantly, although they differ by factors of order unity (and perhaps by factors of $1-\overline{p}$), all of these approaches concur on the basic scaling of Eq. \ref{eq:bp-limit} in the diffusion-limited regime in three dimensions.

The situation is different in one and two dimensions, where Polya's recurrence theorem tells us that, unlike in three dimensions, a random walker is guaranteed eventually to return to its starting point \cite{Redner2001, Bicknell2015}.  Not surprisingly, this recurrence leads to long-time correlations in concentration fluctuations, with the consequences that very long averaging times may be required for meaningful sensor readings in infinite domains \cite{Tkacik2009, Yaron2014} and that the sensing limit in confined domains depends strongly on domain size \cite{Bicknell2015}.

Although they have proven very valuable for understanding many systems, all of these theoretical results (with the notable exception of \cite{Bicknell2015}, which is limited to situations of very high symmetry) are limited to small sensors suspended in an infinite domain of diffusion.  Biological sensors, however, also commonly operate in situations where the infinite domain approximation is not appropriate: either the sensor is not small compared to the volume of the domain where the ligand can diffuse, or the sensor is located near a domain boundary.  This is true, in particular, of many situations where proteins within the cell sense the concentration of other intracellular molecules. Most obviously, cytosolic molecules are often detected by membrane proteins that are, from the perspective of the cytosol, protruding from a reflecting boundary.  To cite one heavily studied example, \textit{E. coli} flagellar motor domains in the inner membrane bias motor rotation in response to cytosolic CheY$^\text{P}$ concentrations \cite{Sarkar2010, Li2011}.  More generally, in both prokaryotes and eukaryotes, proteins often exhibit specific, stereotyped subcellular localizations \cite{Rudner2010, Kumar2002} and thus effectively act as localized sensors for their ligands; unless these protein clusters are both very small and located far from any confining membranes, their sensing precision will differ from the infinite space limit. For example, putative size-sensing proteins in some rod-shaped cells accumulate at mid-cell, forming a cluster whose size is an appreciable fraction of the total cell length \cite{Si2019, Pan2014}. Similarly, the nuclear import of molecules that arrive at the nuclear envelope by diffusion in the cytosol is a crucial step in many biological decisions \cite{Krieghoff2006,deMan2021, Ambrosi2022, Singh2017, Babcock2004}, and the nucleus is typically not small compared to the cell as a whole.  Even within the nucleus, different regulatory elements on the chromosomes can preferentially segregate to the center of the nucleus or to the vicinity of the nuclear envelope \cite{Smith2021, SAKAMOTO2023}, potentially affecting their ability to sense nuclear transcription factor concentrations. Generally speaking, we expect many biological sensing problems to depend on contributions from the sensor's size and location within the enclosed volume of the cell. Beyond the scale of single cells, various mechanisms based on sensing a chemical concentration in a highly confined environment have been proposed to regulate tissue growth and patterning, in which the sensing cells comprise a large fraction of the domain volume and/or are located near the domain boundary \cite{Vollmer2017, PerezMockus2023, Navarro2024, Hufnagel2007, BenZvi2010, BenZvi2011, Vuilleumier2010}. 

Inspired by these examples at the subcellular and the tissue scales, our intention in this work is to begin to understand corrections resulting from sensor confinement to standard results for sensors in infinite domains.  Importantly, we expect that such quantitative corrections will increasingly be measurable and experimentally relevant. Indeed, in many cases it is already possible to obtain large amounts of precise, quantitative, single-cell data at high spatial and temporal resolutions \cite{Jun2018, Micali2016, Skinner2013, Stuart2019}, and analyses that turn on careful quantitative comparisons with physical bounds have already proven fruitful in understanding several systems \cite{Gregor2007, Dubuis2013, Bauer2021, Desponds2020, Brumley2019, Mattingly2021}. In the future, quantitative models that extend beyond simple scaling results, even when differing only by factors of order unity, are likely to be experimentally distinguishable and thus to aid in the interpretation of experimental data in many contexts.  It is thus valuable to figure out when sensor precision in more realistic geometries can be expected to deviate appreciably from the precision in the infinite domain limit.

In the remainder of this paper, we will investigate how the the diffusion-limited sensing precision depends on sensor size and on proximity to reflecting boundaries.  Our results recover the key findings of Bicknell \textit{et al.} \cite{Bicknell2015} for the specific case of sensors centered in a finite-sized domain with complete rotational symmetry, but we explore a considerably wider range of geometries, including situations of lower symmetry and with internal boundaries.  We do not consider receptor binding kinetics but focus on the sensing limit fundamentally set by the physics of diffusion. 

The paper is organized as follows. First, in Sec. \ref{sec:set-up}, we formulate the problem of sensing by a perfect instrument in general terms and recall how the variance in the concentration estimate can be split into contributions from spatial and temporal averaging.  Next, in one dimension (Sec. \ref{sec:1d}), we derive exact expressions for the sensing limit for both small and spatially extended sensors, either centered in the domain or near the domain walls (Fig. \ref{fig:1d} and Table \ref{tab:1d-precision}). Moving to three dimensions (Sec. \ref{sec:3d}), we calculate the corresponding precision limit for sensor configurations in rectangular, cylindrical, and spherical geometries with reflecting boundaries (Figures \ref{fig:rect-cyl} and \ref{fig:spheres}).  In all cases, we find that the precision limit can vary substantially (by as much as factors of 3 or 4) depending on the placement and size of the sensor.

\section{Problem set-up \label{sec:set-up}}
In all the calculations that follow, we assume that $N$ non-interacting molecules diffuse in a domain of volume $V$ so that the concentration of these ligand molecules is $\langle c \rangle = N/V$.  Here, $\langle \ldots \rangle$ represents an ensemble average, and \cavg\ is the average of the fluctuating concentration field $c(\Vec{x},t)$ at every position $\Vec{x}$ and time $t$. An idealized sensor (which could represent, for instance, a cell, a nucleus, or a section of tissue) tries to determine $\langle c \rangle$, but its estimate of this concentration has a nonzero variance $\delta c^2$ because of the stochastic arrival of discrete ligands at the sensor. Our goal is to determine how $\delta c^2$ depends on variables like the sensor size and location.

More specifically, we compute the sensing variance for a ``perfect instrument" in the sense of Berg and Purcell \cite{Berg1977}. In this model, a sensor of volume $v < V$ takes instantaneous counts of the ligand number $n(t)$ within its volume and averages these measurements over a time $T$ to determine the ligand concentration.  (The presence of the sensor does not in any way alter ligand diffusion through the volume $v$.)  As mentioned in the Introduction, the perfect instrument model expression for $\delta c^2$ is expected to agree, up to numerical factors of order unity and perhaps factors that depend on receptor occupancy, with results from more detailed models in the limit that the dominant source of noise is the diffusive arrival of ligand molecules at the sensor \cite{Endres2009, Kaizu2014, RTW2016, Aquino2015}; away from this limit, additional terms may be needed that reflect, for example, the noise associated with stochastic ligand binding and unbinding from receptors, \cite{Bialek2005,Bicknell2015, Berezhkovskii2013, Kaizu2014}. Thus, in particular, we expect that the perfect instrument will give a useful picture of how the physical limit on sensing precision changes when we introduce the effects confinement and domain geometry that are of interest here.

Within the perfect instrument model, the sensor estimates the ensemble average number \avg{n}\ of ligands in its volume as $T^{-1} \int_0^T n(t) dt$.  ($T^{-1} \int_0^T n(t) dt$ is evidently an unbiased estimator of \avg{n}; for our purposes it is not essential to prove whether it is the best estimator in some maximum likelihood or minimum variance sense \cite{Endres2009, Tostevin2009, Govern2012}.)  Because $\avg{n} = \cavg v$, this implies an estimate $(v T)^{-1} \int_0^T n(t) dt$ of the average concentration \cavg.  With the definition $\Delta n(t) \equiv n(t) - \avg{n}$, the variance $\delta n^2$ of the sensor's estimate of $\avg{n}$ can be written as
\begin{align}
\label{eq:deltan2}
\delta n^2 = \frac{1}{T^2} \int_0^T \int_0^T \avg{\Delta n(t) \Delta n(t')} dt\, dt' \; .
\end{align}
The variance $\delta c^2$ of the estimate of \cavg\ then satisfies
\begin{align}
    \frac{\delta c^2}{\langle c \rangle^2} = \frac{\delta n^2}{\langle n \rangle^2} \; .
\end{align}

We can separate the contributions to this variance from spatial averaging and from time averaging by expressing it as \cite{RTW2016, Bicknell2015}
\begin{align}
\label{eq:counting-stats}
    \frac{\delta n^2 }{\langle n \rangle^2} &=  \frac{\delta n _0^2 }{\langle n \rangle^2} \frac{2 \tau }{T}.
\end{align}
In this expression,
\begin{align}
\delta n_0^2 \equiv \avg{\Delta n(t)^2} = \avg{n^2} - \avg{n}^2 
\end{align}
is the variance in a single, instantaneous measurement of the number $n(t)$ of ligands inside the sensor's volume and is set entirely by equilibrium statistical mechanics, independent of any assumptions about the dynamics of particle diffusion. We also introduce the correlation time $\tau$ of $n(t)$  \cite{Berg1977, Bicknell2015},
\begin{align}
\label{eq:tau-def}
    \tau \equiv \frac{1}{\delta n_0^2}\int_0^{\infty} \langle \Delta n(0) \Delta n(t)\rangle dt.
\end{align}
With these definitions, Eq. \ref{eq:counting-stats} is exact in the limit of large $T$; we will always work in this limit.  Eq. \ref{eq:counting-stats} can then be interpreted as saying that the sensor makes $T/(2 \tau)$ independent measurements of the particle number in a time $T$, each with variance $\delta n_0^2$.

When $v \ll V$, the statistics of equilibrium fluctuations are Poissonian, and $\delta n_0^2 = \langle n \rangle$. On the other hand, as $v \rightarrow V$, we expect that $\delta n_0^2 \rightarrow 0$, because the total number $N$ of particles does not fluctuate, and in this limit the sensor counts every particle within the domain.  More generally, $\delta n_0^2/\avg{n}^2$ can be determined from the fact that $n$ measured at a single time must follow a binomial distribution.  Below, we will introduce stochastic dynamics for the field $c(\Vec{x},t)$ that will allow us also to directly compute $\delta n^2/\avg{n}^2$.  We will then be in a position to find $\tau$ from Eq. \ref{eq:counting-stats}, which will provide useful insight into the physical interpretation of many of our results.

\section{\label{sec:1d} precision of sensors in 1D}

\begin{figure}
    \begin{subfigure}{8.6cm}\includegraphics[width=0.999\columnwidth]{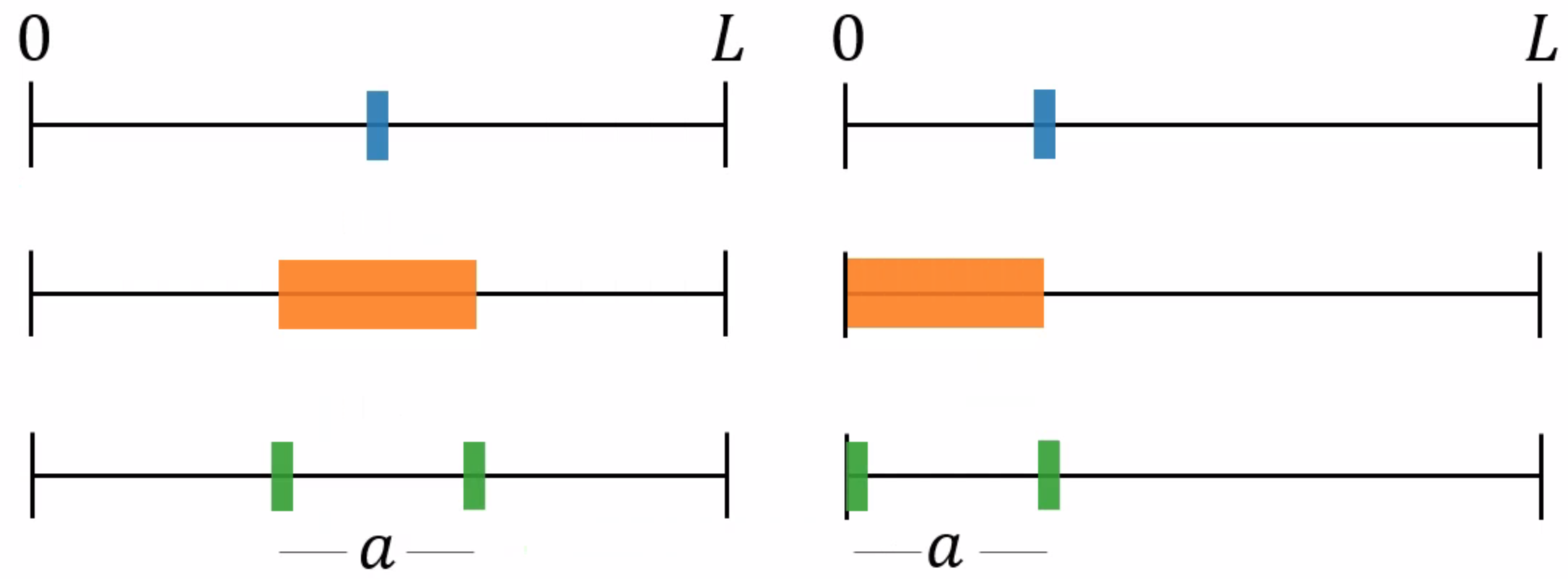}  
        \caption{}
        \label{fig:1d-models}
    \end{subfigure}
    \\
    \begin{subfigure}{8.6cm}
        \includegraphics[width=0.999\columnwidth]{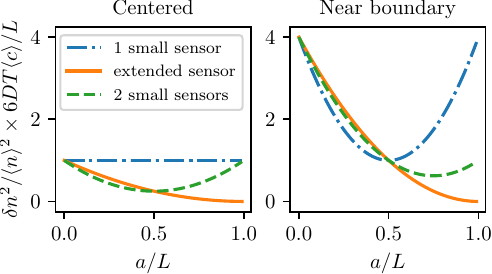}
        \caption{}
        \label{fig:1d-plots}
    \end{subfigure}
     \caption{\label{fig:1d} \justifying Six sensor geometries and their precision limit in 1D. (a) Molecules diffuse freely between reflecting boundaries at $x=0$ and $x=L$. Each of the small sensors (blue and green) has linear size $\ve \rightarrow 0$. Left: sensors centered in the domain; right: sensors near a reflecting boundary. Blue: one small sensor; orange: spatially extended sensor with linear size $a$; green: two small sensors separated by a distance $a$.  In the right column, the blue sensor is a distance $a$ from the boundary, while the orange sensor and the leftmost green sensor contact the boundary. (b) Plot of $\delta n^2 / \langle n \rangle ^2$ for the six sensors as a function of $a/L$ (expressions given in Table \ref{tab:1d-precision}).}
\end{figure}

In this section, we first (Sec. \ref{sec:1d-model}) recall how to add noise to the diffusion equation to model the concentration fluctuations of non-interacting molecules. We next use this equation to calculate the two point correlation function of the ligand concentration.  This correlation function leads directly to an expression for $\delta n^2/\avg{n}^2$, which we then (Sec. \ref{sec:1d-solutions}) evaluate and interpret for the geometries shown in Fig. \ref{fig:1d}(a).

\subsection{Model \label{sec:1d-model}}
In one dimension, we replace the volumes $V$ and $v$ introduced in the preceding section with lengths $L$ and $l$ and the vectorial position $\Vec{x}$ with a scalar $x$. In order to capture the fluctuations in the local concentration $c(x,t)$ caused by the discrete, particulate nature the of the diffusing ligand, we introduce the noisy diffusion equation
\begin{align}
\label{eq:1d-diffusion}
    \pdv{}{t} c(x, t) &= D\pdv[2]{}{x} c(x, t) + \eta(x, t) 
\end{align}
where the Gaussian noise $\eta$ has mean $0$ and correlator
\begin{align}
\label{eq:1d-noise-correlator}
    \langle \eta(x, t) \eta(x', t')\rangle = 2 D \langle c \rangle  \delta(t-t') \pdv{}{x}\pdv{}{x'}\delta(x-x').
\end{align}
Eq. \ref{eq:1d-diffusion}, with noise statistics given by Eq. \ref{eq:1d-noise-correlator}, is a special case of the stochastic Cahn-Hilliard-Cook model, or ``model B" in the study of critical phenomena \cite{COOK1970, LANGER1971, Elliott1989, ChaikinLubensky, Hohenberg1977}. The spatial correlation structure of the noise enforces local mass conservation, while the prefactor $2 D \langle c \rangle$ is chosen so that the fluctuations obey the fluctuation-dissipation theorem \cite{Gardiner2009, ChaikinLubensky}.  Eqs. \ref{eq:1d-diffusion} and \ref{eq:1d-noise-correlator} are expected to give the correct two point correlations of $c(x,t)$, which are all that we require, but with Gaussian noise they of course cannot exactly capture non-Gaussian tails in the distributions of discrete numbers of particles.  It is worth keeping in mind that such deviations from Gaussian behavior are expected to become more pronounced as $\avg{n}$ becomes smaller.

We study Eq. \ref{eq:1d-diffusion} in a domain that extends from $x=0$ to $x=L$ with reflecting boundary conditions
\begin{align}
   0 = \pdv{}{x} c(x, t) \bigg| _ {x=0, L}
\end{align}
and separate $c$ into a constant background and a fluctuating part according to
\begin{align}
    c(x, t) &= \langle c \rangle  + \Delta c(x, t) .
\end{align}
Appendix \ref{app:1d-correlator} solves Eq. \ref{eq:1d-diffusion} for $\langle \Delta c(x, t)  \Delta c (x', t')  \rangle$ by eigenfunction expansion, leading to
\begin{align}
\label{eq:1d-correlator}
    \frac{\langle \Delta c(x, t)  \Delta c (x', t') \rangle}{\langle c \rangle ^2} &=\frac{2}{ \langle c \rangle L}\sum_{p=1}\psi_{p}(x)\psi_{p}(x') e^{-D k^2_{p}|t'-t|}\;\;\;\;\nonumber\\
\end{align}
where $\psi_p$ and $k_p^2$ are the eigenfunctions and eigenvalues that satisfy $(\partial^2 /\partial x^2 )\psi_p = -k_p^2 \psi_p$. Explicitly,
\begin{align}
    \psi_p(x) &\equiv \cos\left(\frac{p \pi x}{L}\right), \;\;\;
    k_p^2 \equiv \frac{p^2 \pi^2}{L^2}.
\end{align}

A sensor with size $l$ has mean ligand occupancy $\langle n \rangle = l \cavg$ and a deviation in occupancy
\begin{align}
\label{eq:dnt-1d}
    \Delta n(t)  \equiv \int_{\text{sensor}}\hspace{-0.6cm} \Delta c(x,t) dx \; ,
\end{align}
where the integral is taken over all points within the sensor, so that $\int_{\text{sensor}} dx = l $.  Combining this expression with Eq. \ref{eq:deltan2} for $\delta n^2$ in terms of $\Delta n$ and the eigenfunction expansion of Eq. \ref{eq:1d-correlator}, we are led to
\begin{widetext}
\begin{align}
\label{eq:dn2-1d}
    \frac{\delta n^2}{\langle n \rangle ^2} &= \frac{1}{l^2 T^2} \int_0^T \int_0^T \iint_{\text{sensor}}\hspace{-0.6cm} \frac{\langle \Delta c(x, t)  \Delta c (x', t') \rangle}{\langle c \rangle ^2} dx dx' dt dt'\nonumber\\
    &= \frac{2}{\langle c \rangle L}\sum_{p=1}\left(\frac{1}{l}\int_{\text{sensor}}\hspace{-0.6cm} \psi_p(x) dx \right)^2\left(\frac{1}{T^2} \int_0^T\int_0^T e^{-Dk_p^2|t-t'|} dt dt'\right) \; .
\end{align}
\end{widetext}
We can directly evaluate the time integral for each mode:
\begin{align}
\label{eq:time-avg}
    \frac{1}{T^2} \int_0^T\int_0^T e^{-Dk_p^2|t-t'|} dt dt' &= \frac{2}{D k_p^2 T} +\frac{2(e^{-Dk_p^2T}-1)}{D^2 k_p^4 T^2}.
\end{align}
As discussed in Sec. \ref{sec:set-up}, we are interested primarily in the limit of large $T$, for which $T D k_p^2 = T D p^2 /(\pi^2 L^2)\gg 1$, so we keep only the first term in Eq. \ref{eq:time-avg}.  This amounts to considering an averaging time $T$ that is much longer than the time $\sim\!L^2/D$ for a molecule to diffuse across the entire domain \cite{Bicknell2015} (but see also Appendix \ref{app:3d-sensor-calc} for important differences between the one-dimensional case introduced here and the three-dimensional case). In this limit, the expression for the fractional variance simplifies to:
\begin{align}
\label{eq:dn-1d}
    \frac{\delta n^2}{\langle n \rangle ^2} &= \frac{4}{\langle c \rangle L D T}\sum_{p=1}\frac{L^2}{p^2 \pi^2}\left(\frac{1}{l}\int_{\text{sensor}}\hspace{-0.6cm} \psi_p(x) dx \right)^2 .
\end{align}

\subsection{1D model solutions \label{sec:1d-solutions}}

To further interpret Eq. \ref{eq:dn-1d}, we need to specify the sensor geometry.  We consider six possibilities, which are illustrated in Fig. \ref{fig:1d}. These geometries differ in the sensor location (centered or near one boundary), the sensor size (``small'' or ``extended''), and the sensor number (one or two).  We define a ``small'' sensor to be one with linear size $\ve \rightarrow 0$, with all results reported to leading order in small $\ve$, and an ``extended'' sensor to be one with finite linear size $a > 0$. 

In Appendix \ref{app:1d-sensor-calc}, we calculate $l^{-1}\int_{\text{sensor}}\psi_p dx$ for each of the six geometries and evaluate the resulting sums to determine $\delta n^2 / \langle n \rangle ^2$ via Eq. \ref{eq:dn-1d}.  The final expressions are listed in Table \ref{tab:1d-precision} and plotted in Fig. \ref{fig:1d}.  (The result for a single, centered small sensor is the same as the diffusion floor for a single receptor reported in \cite{Bicknell2015}.)  Table \ref{tab:1d-precision} also includes the variance $\delta n_0^2/\avg{n}^2$ in a single measurement of $n(t)$ and the correlation time $\tau$ for each geometry. The single measurement variance follows from the fact that, according to equilibrium statistical mechanics, $n(t)$ at a single time must follow a binomial distribution, with probability $l/L$ of being in the sensor and $1-l/L$ of being outside the sensor; here the sensor size $l$ is, respectively, $\ve$, $a$, and $2 \ve$ for one small sensor, one extended sensor, and two small sensors.  We then determine the correlation time using Eq. \ref{eq:counting-stats} and our knowledge of $\delta n^2$ and $\delta n_0^2$.

\begin{table*}
\setlength{\tabcolsep}{1em} 
\begin{center} \begin{tabular}{llll}
\multicolumn{1}{c}{Model} & \multicolumn{1}{c}{$\delta n^2_0/\langle n \rangle ^2$} & \multicolumn{1}{c}{$\tau$} & \multicolumn{1}{c}{$\delta n^2 /\langle n \rangle ^2$} \\
\toprule 
Centered: One small sensor &  $\displaystyle\frac{1}{\langle c \rangle \ve}$ & $\displaystyle\frac{L\ve}{12D}$ & $\displaystyle\frac{L}{6 \langle c \rangle  D T}$ \\
\cmidrule{1-4}
Centered: Extended sensor &  $\displaystyle\frac{1}{\langle c \rangle  a}\left(1-\frac{a}{L}\right)$ & $\displaystyle\frac{La}{12D}\left(1-\frac{a}{L}\right)$ & $\displaystyle\dfrac{L}{6 \langle c \rangle  D T}\left(1-\frac{2a}{L} + \frac{a^2}{L^2}\right)$ \\
\cmidrule{1-4}
Centered: Two small sensors &  $\displaystyle\frac{1}{2\langle c \rangle \ve}$ & $\displaystyle\frac{L\ve}{6D}\left(1-\frac{3a}{L} + \frac{3a^2}{L^2}\right)$ & $\displaystyle\frac{L}{6 \langle c \rangle  D T}\left(1-\frac{3a}{L} + \frac{3a^2}{L^2}\right)$ \\
\cmidrule{1-4}
Boundary: One small sensor &   $\dfrac{1}{\langle c \rangle \ve}$ &$\displaystyle\frac{L\ve}{3D}\left(1-\frac{3a}{L} + \frac{3a^2}{L^2}\right)$&$\displaystyle\frac{2L}{3 \langle c \rangle  D T}\left(1-\frac{3a}{L} + \frac{3a^2}{L^2}\right)$\\
\cmidrule{1-4}
Boundary: Extended sensor &  $\displaystyle\frac{1}{\langle c \rangle  a}\left(1-\frac{a}{L}\right)$ & $\displaystyle\frac{La}{3D}\left(1-\frac{a}{L}\right)$ & $\displaystyle\dfrac{2L}{3 \langle c \rangle  D T}\left(1-\frac{2a}{L} + \frac{a^2}{L^2}\right)$ \\
\cmidrule{1-4}
Boundary: Two small sensors & $\dfrac{1}{2 \langle c \rangle \ve}$& $\displaystyle\frac{2 L\ve}{3D}\left(1-\frac{9a}{4L} + \frac{3a^2}{2 L^2}\right)$ &$\displaystyle\frac{2L}{3 \langle c \rangle  D T}\left(1-\frac{9a}{4L} + \frac{3a^2}{2L^2}\right)$\\
\bottomrule \end{tabular} \end{center}
    \caption{Sensing precision for 1D sensors.  (Results for small sensors are reported to leading order in $\ve$; the different models are defined in Fig. \ref{fig:1d-models} and the accompanying text.)}
    \label{tab:1d-precision}
\end{table*}

Table \ref{tab:1d-precision} and Fig. \ref{fig:1d} reveal several interesting features of the sensing precision.
For example, for a fixed sensor model (one small sensor, two small sensors, or extended sensor) and fixed $a$, we find that $\delta n^2/\avg{n}^2$ for the centered position is always smaller than for the same sensor positioned near the boundary.  Surprisingly, the difference can be as large as a factor of 4. Although $\delta n_0^2/\avg{n}^2$, which is determined entirely by equilibrium considerations, must be the same wherever the sensor is placed, sensors near the boundary have a longer correlation time $\tau$ than centered sensors.  This presumably reflects the fact that molecules can only escape sensors near the boundary on one, rather than two, sides; molecules thus tend to reside longer in the vicinity of the sensor, and correlations in $n(t)$ decay more slowly for boundary sensors.

More unexpected is the relative effectiveness of the different sensor models.  Comparing the centered extended sensor to the centered small sensor, we find that the extended sensor with finite size $a$ has the smaller sensing variance, reflecting the benefit of a larger region of spatial averaging. While this comes at the cost of increasing the correlation time for small $a$, the two effects combine such that larger sensors still perform better than smaller sensors (Table \ref{tab:1d-precision}). However, paradoxically, we find that the extended sensor with size $a$ has \textit{higher} variance than two small sensors separated by $a$, when $a < L/2$, even though the extended sensor samples a larger region of space. The improved spatial averaging offered by the extended sensor comes at the cost of compromising the sensor's ability to make independent measurements and thus of the correlation time $\tau$. The extended sensor eventually lowers its variance below that of the two-sensor model when $a=L/2$, and thereafter the variance decreases to $0$ as $a\rightarrow L$, reflecting the fact that every particle in the domain is being measured in this limit. Proximity to a reflecting boundary further reduces the ability of spatially extended sensors to make independent measurements. Of the three sensing models near the boundary, the small single sensor has the smallest sensing variance for $0<a<L/2$, despite also having the smallest size. Thus, proximity to a boundary can have such a large effect that it is better to give up sensor space near the boundary than to let the part of the sensor near the boundary ``contaminate'' parts farther away.

It is also worth commenting on the scaling of the correlation time with model parameters.  Table \ref{tab:1d-precision} shows that $\tau \sim L \ve/D$ for small sensors, whereas the obvious diffusive timescales in the problem are $\ve^2/D$ for diffusion across the sensor of size $\ve$ and $L^2/D$ for diffusion across the entire domain of size $L$.  How do these combine to give $\tau \sim L \ve/D$?  We argue in Appendix \ref{sec:explain-tau} that, except at very short times, the correlation function $\avg{\Delta n(t) \Delta n(0)}/\delta n_0^2$ can be written in the form $(\ve/L) f(t D/L^2)$ for some function $f$, so that we can think of its integral $\tau$ as coming from the longer timescale weighted by an appropriate prefactor, $\tau \sim (\ve/L) (L^2/D)$.  Thus, although the definition of $\tau$ in Eq. \ref{eq:tau-def} represents a natural notion of a correlation time, its interpretation in systems with multiple competing timescales can be nontrivial; in particular, one should not imagine that the correlation function necessarily decays exponentially on a single timescale $\tau$.

In sum, in this section we have shown that even simple models of one-dimensional concentration sensors can show surprising and strong dependencies on geometry:  Sensors near reflecting boundaries are systematically worse than sensors in the middle of the domain, by as much as a factor of 4 in $\delta n^2/\avg{n}^2$. And strategically placed small sensors can, by virtue of their short correlation times, sometimes outperform larger sensors. 

\section{Precision of sensors in 3D \label{sec:3d}}  
\subsection{Model}
In 3D, the noisy diffusion model is analogous to the 1D version, but with $\partial / \partial x \mapsto \nabla$:
\begin{align}
\label{eq:3d-diffusion}
    \pdv{}{t}c(\Vec{x}, t) &= D\nabla^2 c(\Vec{x}, t) + \eta(\Vec{x}, t),
\end{align}
where $\eta$ has mean $0$ and correlator
\begin{align}
\label{eq:3d-noise-correlator}
    \langle \eta(\Vec{x}, t) \eta(\Vec{x}', t')\rangle = 2 D \langle c \rangle  \delta(t-t') \nabla_{\Vec{x}}\cdot \nabla_{\Vec{x}'}\delta(\Vec{x}-\Vec{x}').\;\;\;\;
\end{align}
As in 1D, we separate $c$ into a uniform background and a fluctuating part, $c(\Vec{x}, t) = \langle c \rangle  + \Delta c(\Vec{x}, t) $. We consider bounded rectangular, cylindrical, and spherical domains, and we solve for the correlator $\langle \Delta c(\Vec{x}, t)  \Delta c(\Vec{x}', t')  \rangle$ in these domains in Appendix \ref{app:3d-correlator}. In general, this correlator takes the form
\begin{align}
    \label{eq:3d-correlator}
    \frac{\langle  \Delta c(\Vec{x}, t)  \Delta c(\Vec{x}', t')  \rangle}{\langle c \rangle ^2} &=\frac{1}{V\langle c \rangle }\sum_{\ell m p}g_{\ell m p}\psi_{\ell m p}(\Vec{x})\psi^*_{\ell m p}(\Vec{x}') \hspace{1.8cm} \nonumber \\
    &\hspace{2.2cm}\times e^{-D k^2_{\ell m p}|t'-t|}, \nonumber\\
\end{align}
where $\psi_{\ell m p}$ and $k_{\ell m p}^2$ are respectively the eigenfunctions and eigenvalues of $\nabla^2$ appropriate to the domain shape, which we tabulate in Appendix Table \ref{tab:3d-geometries}, $g_{\ell m p}$ are normalization constants of the eigenfunctions (Appendix Table \ref{tab:glmp}), and $V$ is the volume of the domain. 

Fluctuations in ligand occupancy within the sensor volume $v$ obey
\begin{align}
\label{eq:dnt-3d}
    \Delta n(t)  \equiv \int_{\text{sensor}}\hspace{-0.6cm} \Delta c(\Vec{x},t)  d^3 x \; ,
\end{align}
which allows us, as in one dimension, to express the time-averaged variance in the occupancy via Eq. \ref{eq:deltan2} as
\begin{widetext}
   \begin{align}
    \label{eq:dn2-3d-T}
    \frac{\delta n^2}{\langle n \rangle ^2} &= \frac{1}{v^2 T^2} \int_0^T \int_0^T \iint_{\text{sensor}}\hspace{-0.6cm} \frac{\langle \Delta c(\Vec{x}, t)  \Delta c (\Vec{x}', t') \rangle}{\langle c \rangle ^2} d^3x d^3 x' dt dt'\nonumber\\
    &= \frac{1}{V\langle c \rangle }\sum_{\ell m p}g_{\ell m p}\left|\frac{1}{v}\int_{\text{sensor}}\hspace{-0.6cm} \psi_{\ell m p}(\Vec{x}) d^3x \right|^2\left(\frac{1}{T^2} \int_0^T\int_0^T e^{-Dk_{\ell m p}^2|t-t'|} dt dt'\right).
\end{align} 
\end{widetext}

As in the one-dimensional case (Eq. \ref{eq:time-avg}), we can evaluate the double integral over time exactly, and the expression simplifies for large $T$.  In this limit,
\begin{align}
    \label{eq:dn2-3d}
    \frac{\delta n^2}{\langle n \rangle ^2} 
    &= \frac{2}{V\langle c \rangle  D T}\sum_{\ell m p}\frac{g_{\ell m p}}{k^2_{\ell m p}}\left|\frac{1}{v}\int_{\text{sensor}}\hspace{-0.6cm} \psi_{\ell m p}(\Vec{x}) d^3x \right|^2.
\end{align}
In Appendix \ref{app:3d-sensor-calc}, we calculate $v^{-1}\int_{\text{sensor}} \psi_{\ell m p}(\Vec{x}) d^3x$ for specific sensor models in rectangular, cylindrical, and spherical domains (illustrated in Figs. \ref{fig:rect-cyl} and \ref{fig:sphere-models}) and evaluate the corresponding sums to obtain $\delta n^2 /\langle n \rangle ^2$.  The Appendix also addresses some subtleties related to how large $T$ must be for the large $T$ limit to apply.  In particular, it shows that, unlike in one dimension, in three dimensions we do not always need $T \gg L^2/D$, where $L$ is a linear dimension of the domain, and thus that we can recover the standard Berg-Purcell formulas for an isolated sensor in an infinite domain. The following section summarizes these results and discusses the behavior of $\delta n^2 / \langle n \rangle ^2$ for the different sensor models.

\subsection{3D model solutions}
\subsubsection{Quasi-1D cylindrical geometry}

\begin{figure}
    \begin{subfigure}{0.49\columnwidth}
        \includegraphics[width=0.99\columnwidth]{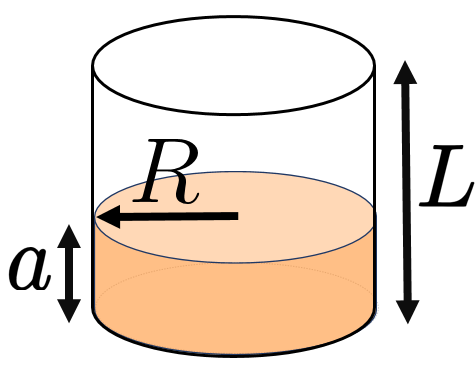}
        \caption{}
        \label{fig:3d-cyl}
    \end{subfigure}
    \hfill
    \centering
    \begin{subfigure}{0.49\columnwidth}
        \includegraphics[width=0.99\columnwidth]{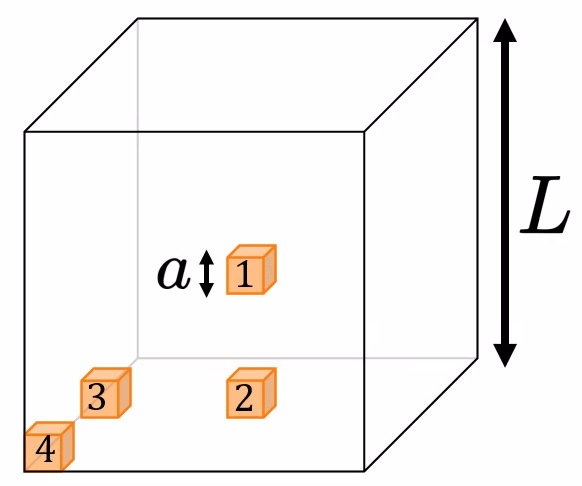}
        \caption{}
        \label{fig:3d-small}
    \end{subfigure}
    \caption{\justifying (a) A cylindrical sensor inside a cylindrical domain; (b) Cubic sensors inside a cubic domain; we consider four different locations (labelled 1--4), with different numbers of faces touching a reflecting boundary, for a small cubic sensor of side $a$.  Black lines outline the domain volume $V$ and orange highlighting indicates the sensor subvolume $v$. The sensing precision for the cylindrical sensor (a) is given by the 1D result (Eq. \ref{eq:quasi-1d-cyl}), while the sensing precision for the four cubic sensors obeys the Berg-Purcell scaling (Eq. \ref{eq:bp-limit}), with prefactors accounting for the proximity of reflecting boundaries (Table \ref{tab:3d-cubes}).}
    \label{fig:rect-cyl}
\end{figure}

We first consider the geometry illustrated in Fig. \ref{fig:3d-cyl}. In this setup, only diffusion parallel to the cylinder axis affects whether or not particles are within the sensor volume; because diffusion in orthogonal directions is independent, the problem then effectively reduces to one dimension, and the sensor's precision only depends on its height $a$ and the domain height $L$. Indeed, in Appendix \ref{app:cyl-calc} we compute $\delta n^2 / \langle n \rangle ^2$ from $\langle \Delta c(\Vec{x}, t) \Delta c(\Vec{x}', t')  \rangle / \langle c \rangle ^2$ expressed in cylindrical coordinates and find that
\begin{align}
\label{eq:quasi-1d-cyl}
    \frac{\delta n^2}{\langle n \rangle ^2}   &= \frac{2L}{3DT(\pi R^2\langle c \rangle )}\left(1-\frac{2a}{L} + \frac{a^2}{L^2}\right) \; .
\end{align}
This is the same as the 1D result for a sensor near a reflecting boundary (Table \ref{tab:1d-precision}) after replacing the 1D concentration with the 3D concentration times the radial and angular surface integral: $\langle c \rangle  \mapsto \pi R^2 \langle c \rangle $. 

\subsubsection{Example: Toy model of size sensing in the \protect\rm{Drosophila} eye disc}
\label{sec:drosophila-estimates}
As an illustrative application of our cylindrical sensor model, we consider a simple description of size-sensing by a developing tissue. In this picture, which has been proposed to explain size control in the developing \textit{Drosophila melanogaster} eye disc (the tissue that gives rise to the adult eye), a fixed amount of a rapidly diffusing cytokine (the protein Unpaired, Upd) is expressed early in development and then is uniformly diluted as the tissue grows, with growth arresting when its concentration falls below a threshold \cite{Vollmer2017}.  Errors in sensing the Upd concentration could thus contribute to errors in adult eye size, making it useful to understand how well realistic concentrations can be detected. Other mechanisms of proliferation arrest have also been implicated in eye disc size control \cite{Navarro2024, Wartlick2014}, and thus the Upd-based dilution model likely oversimplifies the full size control mechanism. Nonetheless, we will consider this dilution model here for its illustrative value.

The size of the adult \textit{Drosophila} eye, scaled by body size, varies at the level of about $1\%$ \cite{Navarro2024}. Here we ask whether this level of precision is plausibly consistent with noise in the measurement of Upd concentration. We approximate the shape of the eye disc as a cylinder, as in Fig. \ref{fig:3d-cyl}. The bottom surface of the cylinder represents the reflecting, apical surface of the disc epithelium proper, which is composed of epithelial cell membranes joined by adherens and septate junctions. The top surface of the cylinder corresponds to the peripodium, and these two opposing membranes, separated by a distance $L$, enclose a luminal space in which Upd is confined to diffuse \cite{Vollmer2017}. The orange volume of height $a$ corresponds to a volume near the apical surface in which cells are sensitive to Upd molecules diffusing in the lumen. We do not consider any downstream cellular communication between the cells of the epithelium \cite{Fancher2017}, but focus only on the limit set by diffusion of Upd, assuming the tissue acts as a ``perfect instrument" which counts every Upd molecule. 

We use Eq. \ref{eq:quasi-1d-cyl} and conservative parameter estimates to ask whether the integration time required to achieve $1\%$ precision in the measurement of Upd concentration is consistent with the time scales associated with growth of the disc. The diffusion constant of Upd is $D \approx \SI{0.7}{\micro\meter^2\second^{-1}}$, the disc area is $\pi R^2 \approx \SI{6e4}{\micro\meter^2}$, and the height of the lumen is $L\approx \SI{5}{\micro\meter}$ \cite{Vollmer2017}. In the absence of a direct measurement of $a$, we will conservatively consider the limit $a\rightarrow 0$. The mean Upd concentration is not known precisely but is bounded by $\cavg \gtrsim \SI{0.3}{\micro\meter^{-3}}$ \cite{Wright2011}. The growth rate of the disc is roughly $\sim \SI{0.1}{\micro\meter^2\second^{-1}}$ \cite{Vollmer2017}, which is appreciably less than $D$, and we therefore assume that the Upd concentration is equilibriated across the disc as the disc grows. Using these parameter values and Eq. \ref{eq:quasi-1d-cyl}, we conclude that the time $T$ required to measure $\cavg$ with 1\% precision is $T \sim \SI{3}{\second}$. However, this time is less than the time for UpD to diffuse a distance $L$, $L^2 / 2D \sim \SI{20}{\second}$, below which Eq. \ref{eq:quasi-1d-cyl} is not valid. Allowing for a longer integration time, a precision of $0.1\%$ can be achieved with an averaging time $T\sim \SI{300}{\second}$. $\SI{300}{\second}$ is much less than the duration of the last phase of disc growth, which is on the order of at least $\sim \SI{30}{\hour}$ \cite{Vollmer2017}. Presumably, the decision to stop growth is made on a shorter time scale than this, on the order of hours. Since this time scale is much greater than our estimate (itself achieved with conservative bounds on the parameter values) $T\sim \SI{300}{\second}$ to achieve $0.1\%$ precision in the concentration measurement, which is already an order of magnitude lower than the amount of variability measured in adult eyes, our estimate suggests that errors in sensing Upd concentration are not likely to be important contributors to developmental variability in the size of the \textit{Drosophila} eye disc. Rather, it seems likely that other factors, which could include anything from errors in the amount of Upd that is initially expressed to noise in downstream communications between cells, are predominant.

\subsubsection{Small, cubic sensors near boundaries}

\begin{table}
    \begin{center}
        \begin{tabular}{ll}
            \multicolumn{1}{c}{(Label) Sensor position \;\;\;\;\;\;}&  
            \multicolumn{1}{c}{$ \delta n^2/\langle n \rangle ^2$} \\
            \toprule 
            (1) Center of domain& $2.32 \displaystyle \times \frac{4}{\pi^3 \langle c \rangle  D Ta}$\\
            \cmidrule{1-2}
            (2) Center of one face & $3.53\displaystyle \times \frac{4}{\pi^3 \langle c \rangle  D T a}$ \\
            \cmidrule{1-2}
            (3) Center of one edge & $5.61 \displaystyle \times \frac{4}{\pi^3 \langle c \rangle  D T a}$\\
            \cmidrule{1-2}
            (4) Corner of domain & $9.28 \displaystyle \times \frac{4}{\pi^3 \langle c \rangle  D T a}$\\
            \bottomrule
        \end{tabular}
    \end{center}
    \caption{\justifying Sensing precision for small cubic sensors of side $a$, to leading order in small $a$.  Numbers correspond to the labelled sensor locations in Fig. \ref{fig:3d-small}. The presence of reflecting boundaries increases the sensing variance by an increasing prefactor, up to a factor of 4 between the sensor in free space (sensor 1) and the sensor in the corner (sensor 4). }
    \label{tab:3d-cubes}
\end{table}

We next investigate how boundaries affect sensing precision in a truly three-dimensional situation.  To this end, we consider 4 cubic sensors with linear size $a$ in a cubic domain of side $L$, as illustrated in Fig. \ref{fig:3d-small}.  In contrast to elsehwere in this paper, for the cubic sensors we confine ourselves to the limit $a \ll L$.  (Although elsewhere we refer to infinitesmal lengths as $\ve$, here we prefer to retain the name $a$, to simplify comparison with results for spherical sensors in the next section.) For sensors much smaller than the domain size, the standard scaling argument by Berg and Purcell does not change with proximity to a boundary, since the boundary does not introduce another length scale until the sensor's size is of the same order as the domain size. We therefore expect that each sensor should have a sensing precision which scales like $a^{-1}$, with proximity to a reflecting boundary changing the sensing precision by a numerical prefactor. To investigate this, in Appendix \ref{app:small-rect-calc} we calculate this prefactor for each of the four small cubic sensors and report our results in Table \ref{tab:3d-cubes}. We find that, compared to the sensor in free space, the sensor in a corner of the domain has a sensing variance 4 times larger.  The sensor near one boundary and the sensor near two boundaries have prefactors intermediate between these two cases. As was the case in one dimension, proximity to a reflecting boundary hence makes it more difficult for the diffusing molecules to enter or leave the sensor volume and so substantially increases the correlation time $\tau$. 

\subsubsection{Spherical sensors}

\begin{figure}
    \centering
    \begin{subfigure}{8.6cm}
        \centering
        \includegraphics[width=8.6cm]{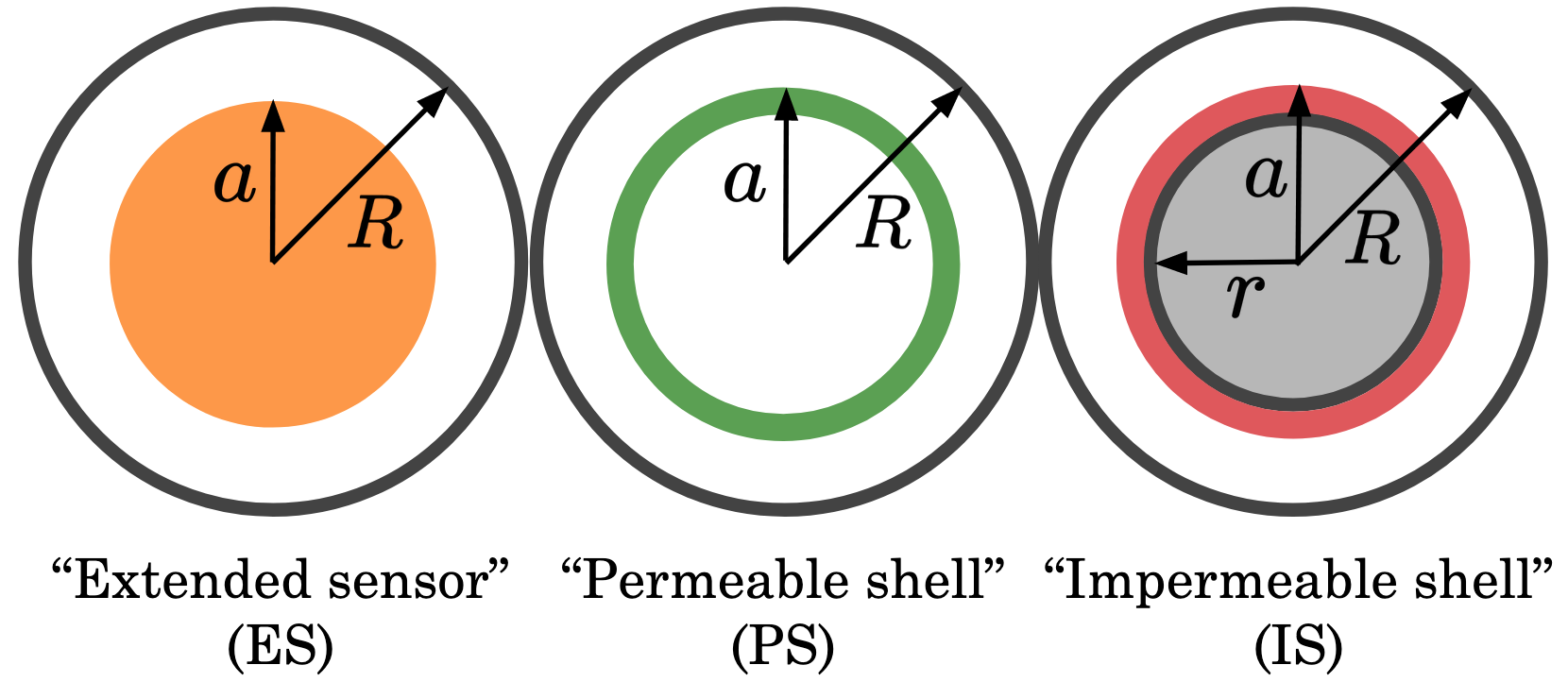}
    \caption{}
    \label{fig:sphere-models}
    \end{subfigure}
    \\
    \begin{subfigure}{8.6cm}
        \includegraphics[width=8.6cm]{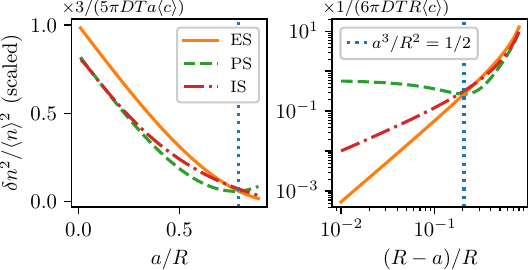}
    \caption{}
    \label{fig:sphere-plots}
    \end{subfigure}
    \caption{\justifying Spherical sensor models and their sensing precision. (a) Cross-sections of the models, in which the colored regions illustrate the sensing volume. In all three models, the sensor has radius $a$ and is concentric with a spherical domain of outer radius $R$. The ES senses particles within a permeable, spherical volume of radius $a$. The PS senses particles within a thin spherical shell of radius $a$ and width $\ve \ll a$. The IS has a second reflecting boundary at $\rho = r$ (where $\rho$ is the radius in spherical coordinates), and ligand diffuses only between $\rho=r$ and $\rho=R$. The sensing shell has width $\ve = a - r \ll a$. (b) Sensing precision for the three models as a function of sensor size (expressions given in Table \ref{tab:3d-scalings}). The three models have sensing precisions with a similar magnitude when $a^3 \lesssim R^3/2$ (left, vertical blue dotted line) but have different asymptotic behavior when $a^3 \gtrsim R^3/2$ (right).  Note that the variance $\delta n^2/\avg{n}^2$ is non-dimensionalized by different factors, listed above the graphs, in the two panels.}
    \label{fig:spheres}
\end{figure}

\begin{table*}
\setlength{\tabcolsep}{1.2em}
\begin{center} \begin{tabular}{llll}
\multicolumn{1}{l}{Model} & \multicolumn{1}{c}{\makecell[c]{$\;\;\delta n^2_0 /\langle n \rangle ^2$}} & \multicolumn{1}{c}{\makecell[c]{$\tau$  }} & \multicolumn{1}{c}{\makecell[c]{$\delta n^2/\langle n \rangle ^2$}} \\[0.0cm]
\toprule 
ES & $\displaystyle \frac{1}{\frac{4\pi}{3}a^3 \langle c \rangle } \left(1-\frac{a^3}{R^3}\right)$ & $\displaystyle \frac{2a^2}{5D}\left(1 - \frac{3a}{2R} + \frac{a^3}{2R^2}\right)\left(1-\frac{a^3}{R^3}\right)^{-1}$ & $\displaystyle \frac{3}{5 \pi  \langle c \rangle  D T a} \left(1-\frac{3 a}{2 R}+\frac{a^3}{2 R^3}\right)$ \\
\cmidrule{1-4}
PS & $\displaystyle \frac{1}{4\pi \langle c \rangle  \ve a^2}$ & $\displaystyle \frac{\ve a}{D}\left(1-\frac{9a}{5R}+\frac{a^3}{R^3}\right)$&   $\displaystyle \frac{1}{2\pi \langle c \rangle DTa}\left(1-\frac{9a}{5R}+\frac{a^3}{R^3}\right)$\\
\cmidrule{1-4}
\multicolumn{1}{l}{IS} &$\displaystyle \frac{1}{4\pi \langle c \rangle  \ve a^2}$ & \makecell[l]{$\displaystyle \frac{\ve a}{D}$ \hspace{1.2cm}(small $a$)\\[0.4cm]
$\displaystyle \frac{\ve}{3D}(R-a)$ \,\,(small $R-a$)} &\makecell[l]{$\displaystyle \frac{1}{2\pi \langle c \rangle DTa}$ \,\,\,\,(small $a$)\\[0.4cm]
$\displaystyle \frac{(R-a)}{6\pi \langle c \rangle  D T a^2}$ \,\,(small $R-a$)} \\ 
\bottomrule \end{tabular} \end{center}
    \caption{\justifying Sensing precision for 3 spherical sensor models in 3D (Fig. \ref{fig:spheres}). The expressions for $\delta n^2 / \langle n \rangle^2$ for the ES and IS hold for arbitrary $a/R$, while those for the IS are given to leading order in small $a$ and in small $R-a$. All results are reported to leading order in small $\ve$.}
    \label{tab:3d-scalings}
\end{table*}

Finally, we consider 3 spherical sensor models (Fig. \ref{fig:sphere-models}).  In all 3 cases, the sensor has spherical symmetry and is concentric with the spherical domain of outer radius $R$.  The spatially extended sensor (ES) consists of a spherical sensor volume of radius $a$.  The permeable sensor (PS) is a thin, permeable shell of radius $a$ and thickness $\ve \ll a$.  Unlike the other two geometries, the impermeable sensor (IS) has an impermeable, reflecting inner sphere of radius $r$ in addition to the outer sphere of radius $R$; the sensor then consists of a spherical shell of thickness $\ve \ll a$ extending from $\rho = r$ to $\rho = a = r + \ve$, where $\rho$ is the radial coordinate.  For the PS and IS, we report results to leading order in small $\ve$.  The details of the calculation of $\delta n^2/\avg{n}^2$ for the three models are presented in Appendix \ref{app:sphere-calcs}, and the results are summarized in Table \ref{tab:3d-scalings} and Fig. \ref{fig:sphere-plots}.

We first compare the models in the small $a$ limit. As in 1D, we find that the sensor with the largest sensing volume, the ES, paradoxically has the \textit{largest} sensing variance (although the effect is not as strong as in 1D).  The ES's weaker performance can again be traced to a long correlation time that overwhelms its advantage in $\delta n_0^2/\avg{n}^2$. Unlike in 1D, this effect holds at leading order in $a/R$; this difference may reflect the fact that $\delta n^2/\avg{n}^2$ depends on the sensor size at leading order in three but not in one dimension. (Encouragingly, our expression for the PS is exactly the same as the diffusion floor for a ring of receptors obtained in Ref. \cite{Bicknell2015} (for arbitrary $a/R$) and in Ref. \cite{Berezhkovskii2013} (for $a/R \rightarrow 0$).)

Strikingly, just as in 1D, the correlation time $\tau$ of the ``small" sensors (i.e. the PS and the IS) scales like $\ve a/D$ for small $a$ (see Sec. \ref{sec:1d-solutions} and Appendix \ref{sec:explain-tau}).  It is very tempting to speculate that this scaling has a similar origin, with most particles that are present at $t=0$ leaving the sensor in times of order $\ve^2/D$, and the integral of the correlation function $\avg{n(t)n(0)}/\delta n_0^2$ dominated by its decay for longer times, where one might imagine it can be written in the form $(\ve/a) f(D t/a^2)$.


It is also surprising that the PS and the IS have the same sensing precision in the small $a$ limit. We might have expected, considering the results from the rectangular domain, that proximity of a small sensor to a boundary, as in the IS, would increase the correlation time and thus the sensing variance.  There are, however, differences between the spherical and cubic geometries that can help to rationalize the comparatively strong performance of the IS.  If we imagine introducing a boundary by moving from cubic sensor 1 to cubic sensor 2 (Fig. \ref{fig:3d-small}), it seems natural that this should increase the correlation time $\tau$:  Whereas a molecule can leave the sensor through six faces in case 1, it has only 5 faces available in case 2 and so should spend longer in the sensor.  Moreover, once the particle has left the sensor, the presence of the reflecting boundary can only close off avenues for it to escape to infinity and thus increase its probability of returning to the sensor.  The corresponding arguments are murkier if we pass from the PS to the IS by introducing a spherical boundary of radius $r$.  The addition of the boundary evidently makes it harder for molecules to exit the sensor and thus should prolong their residence time and so increase the correlation function $\avg{n(t)n(0)}/\delta n_0^2$ at short times.  On the other hand, molecules that leave the PS towards the inside of the sphere must eventually pass through the sensor again before they can escape to infinity, enhancing correlations on longer timescales of order $a^2/D$.  The presence of the reflecting boundary in the IS closes off this avenue for recurrence; once particles do manage to leave the sensor, they do so towards the outside of the sphere and have a finite probability never to return.  Thus, it seems plausible that the PS may actually have larger correlations than the IS at long times.  How these two countervailing effects combine to determine which correlation time is longer is not obvious, though at the moment it appears that the exact equality between the IS and the PS in the small sensor limit may be coincidental.

Turning away from the small $a$ limit, the three sensor models have a similar sensing precision for $a^3 \lesssim R^3/2$ ($a \lesssim 0.8 R$).  When $a^3 = R^3/2$, the ES senses with the same variance as the PS, and for larger $a$ it performs better than the PS. This situation is similar to 1D, in which the two-sensor model is roughly a 1D ``shell", since it is comprised of the set of points equidistant from the origin, and is thus analogous to the PS. In the Introduction, we noted that one biological situation in which corrections to the Berg-Purcell scaling may be important is when the nucleus acts as a subcellular sensor of cytosolic concentrations. If the cell nucleus occupies about $7\%$ of the total cell volume \cite{Jorgensen2007}, $a/R \approx 0.4$, and the linear corrections to the Berg-Purcell scaling (listed in Table \ref{tab:3d-scalings}) due to confinement are already significant. In the large sensor limit, the differences among the precisions of the different sensors are even more pronounced, and each sensor model has different asymptotic behavior (Fig. \ref{fig:sphere-plots}). For cells with very large nuclei (for instance, lung cancer cells can have typical $a/R \gtrsim 0.85$ \cite{Vollmer1982}), models of nuclear sensing therefore need to take careful account of the specific sensor model.

\section{Discussion}
Many biological systems measure chemical concentrations within confined domains; thus, it is valuable to understand how the physical limits of chemosensation depend on domain and sensor geometry. For instance, as mentioned in the Introduction, biological decisions can depend quantitatively on the nuclear import of molecules that arrive at the nucleus by diffusion or on the binding of cytosolic molecules to membrane-anchored proteins. Beyond the scale of single cells, tissue growth and patterning often depend on measurements of concentrations in highly confined developmental environments. 

To lay the theoretical groundwork to understand such situations quantitatively, in this work we have investigated how sensor size, shape, and placement in confining domains lead to
corrections to established expressions for sensor precision in infinite domains.  Our main analytic results are summarized in Tables \ref{tab:1d-precision}--\ref{tab:3d-scalings} and plotted in Figs. \ref{fig:1d-plots} and \ref{fig:sphere-plots}.   In general, we find that confinement can have nontrivial quantitative effects on sensing precision.  For example, a sensor near a reflecting boundary can have $\delta n^2/\avg{n}^2$ as much as a factor of 4 larger than a sensor of the same size far from boundaries. Similarly, $\delta n^2/\avg{n}^2$ drops when the sensor begins to occupy an appreciable fraction of the domain where the ligand can diffuse; for example, in one dimension, a sensor that occupies half the domain has one quarter the scaled variance of a very small sensor. More remarkably, we also uncovered situations, in both one and three dimensions, where smaller sensors can be more effective than larger sensors.  This behavior arises because the smaller sensor's shorter correlation time $\tau$ more than compensates for its larger single measurement variance $\delta n_0^2/\avg{n}^2$. The effect seems in particular to occur in cases where the smaller sensor sits at the boundary of the larger sensor, consistent with the intuition that all of the information about $\avg{c}$ is in fact conveyed by the initial arrival of new particles, not by their subsequent diffusion in and around the sensor \cite{Berg1977,Endres2009,Aquino2015}.

More specifically, in 1D we considered six different sensor configurations, chosen to illustrate how the sensing precision depends on the sensor position and size (Fig. \ref{fig:1d}). We derived exact expressions for $\delta n^2/\avg{n}^2$ for these six configurations (Table \ref{tab:1d-precision}), several of which exhibit non-monotonic dependencies on sensor size and position.
In 3D, we quantified how the prefactor multiplying the basic Berg-Purcell scaling (Eq. \ref{eq:bp-limit}) varies with proximity to reflecting boundaries for small sensors (Fig. \ref{fig:3d-small} and Table \ref{tab:3d-cubes}). We analyzed a cylindrical sensor (Fig. \ref{fig:3d-cyl}), showing explicitly how this specific 3D sensing problem reduces to a quasi-1D problem, and we used this sensor model as an illustrative toy model for size-sensing in growing \textit{Drosophila} eye discs (Sec. \ref{sec:drosophila-estimates}). Lastly, we considered three models of spherical sensors of radius $a$ in a spherical domain of radius $R$ (Fig. \ref{fig:sphere-models} and Table \ref{tab:3d-scalings}). For one of these models, the extended sensor (ES), we extended a result of Berg and Purcell \cite{Berg1977} for $a \ll R$ to arbitrary $a/R$. We found that the expressions for our spherical sensor models already deviate from each other at linear order in $a/R$ and have significantly different asymptotic behavior as $a \rightarrow R$ (Fig. \ref{fig:sphere-plots}). 

An important open question is how exactly our results on the perfect instrument model would change with a more detailed and biologically realistic description.  A comparison with the findings of Bicknell \textit{et al.} \cite{Bicknell2015}, who used the FDT and mass-action kinetics (as in \cite{Bialek2005}) to find the sensing precision with explicit binding to receptors, is instructive here:  Our expression for $\delta n^2/\avg{n}^2 = \delta c^2/\avg{c}^2$ for the small, centered sensor in 1D matches their diffusion-limited sensing bound for the same geometry; similarly, in 3D, our permeable sensor (PS) coincides exactly, for arbitrary $a/R$, with the diffusion limit for their spherical shell of receptors.  It is not difficult to understand the mathematical origin of these correspondences; the two calculations have essentially the same structure, leading to the same sums over eigenfunctions and eigenvalues of the Laplacian in a given geometry.  We thus hypothesize that, in general, our perfect instrument results give the diffusion limit of a Bicknell-style FDT calculation of sensing precision for receptors spread across the sensor volume.  We might then expect that receptors that actively internalize or degrade ligands would have better performance by a numerical factor \cite{Endres2009,Aquino2015}. For example, compared to a spherical perfect absorber of the same radius in infinite space (i.e. $a \ll R$), our PS and IS models have a larger variance by a factor of 2 \cite{Endres2008} (Fig. \ref{fig:sphere-models} and Table \ref{tab:3d-scalings}), and the same factor of 2 appears to hold for a circular absorber in a finite-sized two-dimensional domain \cite{Bicknell2015}, despite the different recurrence properties of diffusion in lower dimensions.

It is somewhat less clear what additional modifications would be needed to capture non-mean-field effects (absent from mass-action kinetics and thus from the FDT approach of \cite{Bicknell2015}) caused by correlations from repeated unbinding and rebinding of the same, discrete ligand to a receptor.  For a single receptor in an infinite 3D domain, analytic arguments and simulations concur that an additional dependence on the average receptor occupancy is introduced \cite{Berezhkovskii2013,Kaizu2014}. On the other hand, Brownian dynamics simulations of a single receptor in a confined, one-dimensional domain in \cite{Bicknell2015} indicate that, in this case, the mass-action/FDT approach is essentially correct.  The resolution of this apparent discrepancy is beyond the scope of the current paper, though it is worth noting that the two conclusions differ in the order of limits that they implicitly assume: The 3D result uses an infinite domain and allows ligands to permanently escape the vicinity of the receptor, whereas the 1D result averages over a time $T$ much larger than the time $L^2/D$ to diffuse across the domain of size $L$, so that all ligands pass near the receptor many times.  Further complications are introduced when correlations among multiple receptors are considered, though some evidence suggests that mean-field-like behavior is recovered in the limit of many receptors \cite{Berezhkovskii2013,Kaizu2014}.

The most immediate application of our results will likely be to biological systems that operate near the physical limits of chemosensation, where the effects considered here could become important.  One might speculate, for example, that cells or tissues could sometimes place sensors away from boundaries in order to improve their ability to estimate concentrations.  In the future, it will also be interesting to extend our work to investigate limits on gradient sensing \cite{Endres2008} and cell-cell communication \cite{Fancher2017} in confined domains.  More broadly, our results provide a starting point to understand when geometric confinement must be taken into account in problems where the physics of diffusion fundamentally limits how precisely information can be transmitted.

\begin{acknowledgments}
We acknowledge funding from HFSP grant RGP0031/2020, NSF award number DMR-2243624, and Simons Fellow grant 919564 from the Simons Foundation.
\end{acknowledgments}

\appendix
\section{Correlator in one dimension}
\label{app:1d-correlator}
In this appendix we will derive the correlator of concentration fluctuations $\langle \Delta c(x, t) \Delta c(x', t)\rangle$ in 1D. In addition to deriving the results in 1D, this appendix will also serve as a simple and general outline of our method and nonmenclature, which will carry over into the 3D derivation (Appendix \ref{app:3d-correlator}). The 3D method involves more complicated orthogonality and normalization conditions of the eigenfunctions, but the logic of the derivation is the same.

In 1D, we consider the Langevin-diffusion equation (main text Eq. \ref{eq:1d-diffusion}):
\begin{align}
    \pdv{}{t} c(x, t) &= D\pdv[2]{}{x} c(x, t) + \eta(x, t)
\end{align}
where the Gaussian noise $\eta$ has the statistical properties
\begin{align}
    \langle \eta (x, t) \rangle &= 0 \\
    \langle \eta (x, t) \eta(x', t')\rangle &= 2 D \langle c \rangle  \delta(t-t') \pdv{}{x}\pdv{}{x'}\delta (x-x').\nonumber\\
\label{eq:noise-correlator}
\end{align}
The concentration field $c$ is composed of a constant background $\langle c \rangle $ plus fluctuations, i.e.
\begin{align}
    c(x, t) &= \langle c \rangle  + \Delta c(x, t) .
\end{align}
$\Delta c$ thus obeys the Langevin-diffusion equation. We consider a bounded domain with reflecting boundaries at $x=0$ and $x=L$, and we consider steady-state fluctuations in which memory of the initial condition has been erased, i.e. we consider
\begin{align}
\label{eq:dc-1d-ld}
   \pdv{}{t} \Delta c(x, t)  - D\pdv[2]{}{x} \Delta c(x, t)  &= \eta(x, t)\nonumber\\ 
    \pdv{\,\Delta c}{x}\Big \vert_{x=(0,L)} &= 0\nonumber\\
    \Delta c(x, -\infty) &= 0.
\end{align}

We solve the differential equation for $\Delta c$ by eigenfunction expansion; we consider the eigenfunctions of $\partial^2 / \partial x^2$, which satisfy
\begin{align}
    \pdv[2]{}{x}\psi_p(x) &= -k_p^2 \psi_p(x).
\end{align}
The eigenfunctions which satisfy the above, as well as the no-flux boundary conditions, are the cosine eigenfunctions
\begin{align}
    &\psi_{p}(x) =  \cos{\left(\frac{p\pi x}{L}\right)}
\end{align}
($p = 0, 1, 2, \ldots$) with corresponding eigenvalues
\begin{align}
    k_p^2 = \frac{\pi^2 p^2}{L^2}.
\end{align}
We then expand $\Delta c$ in the cosine eigenbasis with time-dependent coefficients $a_p(t)$:
\begin{align}
    \Delta c = \sum_{p=0}^{\infty}a_p(t)\psi_p(x).
\end{align}
In order to solve for the $a_p$, we substitute the expansion in the diffusion equation:
\begin{align}
\label{eq:1d-coeffs}
    \sum_{p}\left( \dot{a}_{p} \psi_{p}-D\pdv[2]{}{x} \psi_{p}a_p\right)&= \eta\nonumber\\
    \sum_{p}\left( \dot{a}_{p}+D k_{p}^2 a_p\right)\psi_{p}&= \eta
\end{align}
and we make use of the orthogonality condition of the eigenfunctions in order to find the coefficients. The orthogonality condition is
\begin{align}
\label{eq:cosine-orthogonality-1d}
    \int_0^L \cos\left(\frac{p \pi x}{L}\right)\cos\left(\frac{p' \pi x}{L}\right) dx &= \begin{cases}
    \frac{L}{2} & p = p' > 0\\
    L & p = p' = 0\\
    0 & p\neq p'
    \end{cases}\nonumber\\
    &\equiv L \delta_{pp'} h_p \nonumber\\
\end{align}
where the last line defines
\begin{align}
    h_p \equiv \begin{cases}
        \frac{1}{2} & p>0\\
        1 & p=0
    \end{cases}.
\end{align}
Multiplying both sides of Eq. \ref{eq:1d-coeffs} by $\psi_{p'}$, using the dummy integration variable $r$ in place of $x$, integrating in $r$ from $0$ to $L$, and re-indexing $p'\mapsto p$, the differential equations for the coefficients read:
\begin{align}
\sum_{p=0}(&\dot{a}_p + Dk_p^2 a_p) \int_0^L \psi_p (r)\psi_{p'}(r) dr =\int_0^L \psi_{p'}(r) \eta(r, t) dr\nonumber\\
    &\dot{a}_p(t) + D k_p^2 a_p(t) =\frac{1}{L h_p}\int_0^L \psi_p(r) \eta(r, t) dr.
\end{align}
With each mode subject to the initial condition
\begin{align}
    a_p(-\infty) &= 0,
\end{align}
the solutions for the $a_p$ are
\begin{align}
    a_p(t) = \frac{1}{L h_p} \int_{-\infty}^{t}e^{-Dk_p^2(t-s)}\int_0^L \eta(r, s)\psi_p(r) dr ds.\nonumber\\
\end{align}
We note that, for the $p=0$ mode, $\psi_0$ is a constant, and the integral of $\eta$ representing conserved fluctuations over the entire domain is zero; therefore $a_0=0$. We introduce the coefficients $g_p$, whose generalization will prove especially useful in our 3D calculation, in order to explicitly track this zero contribution for the zero mode. We define $g_p$: 
\begin{align}
    g_p &\equiv \begin{cases}
        0 & p=0\\
        \dfrac{1}{h_p} & \text{otherwise}
    \end{cases}\nonumber\\
    &= \begin{cases}
        0 & p=0\\
        2 & p>0
    \end{cases}.
\end{align}
We write the expansion for $\Delta c(x, t) $:
\begin{align}
    \Delta c = \frac{1}{L}\sum_{p=0}^{\infty} &g_p \psi_p(x)\times \nonumber \\ &\int_{-\infty}^{t}e^{-Dk_p^2(t-s)}\int_0^L \eta(r, s)\psi_p(r) dr ds\nonumber\\
\end{align}
with $r$ and $s$ dummy variables of spatial and time integration, respectively. To derive the correlator $\langle \Delta c (x, t)\Delta c(x', t')\rangle$, we write
\begin{widetext}

\begin{align}
\label{eq:correlation-1d}
    \langle \Delta c(x, t) \Delta c(x', t')\rangle =  \frac{1}{L^2}\sum_{p=0}\sum_{p'=0}&g_p g_{p'}\psi_{p}(x)\psi_{p'}(x') \int_{-\infty}^t\int_{-\infty}^{t'}e^{-D k^2_{p}(t-s)}e^{-D k^2_{p'}(t'-s')}\nonumber\\
    & \times \int_{0}^L\int_{0}^L\psi_{p}(r)\psi_{p'}(r')\langle \eta(r, s), \eta(r', s') \rangle drdr'dsds'.
\end{align}
We will first evaluate the double spatial integral, so we substitute the expression for the conserved noise correlator (Eq. \ref{eq:noise-correlator}):
\begin{align}
\label{eq:double-time-integral}
    \int_{0}^L\int_{0}^L\psi_{p}(r)\psi_{p'}(r')\langle \eta(r, s), \eta(r', s') \rangle drdr'&=2D\langle c \rangle  \delta(s-s')\int_0^L\int_0^L\psi_{p}(r)\psi_{p'}(r') \pdv{}{r}\pdv{}{r'}\delta(r-r') drdr'\nonumber\\
    &= -2D\langle c \rangle  \delta(s-s')\int_0^L \psi_{p}(r) \pdv[2]{}{r}\psi_{p'}(r) dr\nonumber
    \\
    &= 2D\langle c \rangle  \delta(s-s')k^2_{p'}\int_{0}^L\psi_{p}(r) \psi_{p'}(r) dr.
\end{align}
We evaluate the spatial integral as in Eq. \ref{eq:cosine-orthogonality-1d}:
\begin{align}
    \int_0^L \psi_{p}(r) \psi_{p'}(r) dr &= L\delta_{pp'}h_p.\nonumber
\end{align}
Substituting into Eq. \ref{eq:correlation-1d}, combining constant prefactors, collapsing the $p'$ sum and noting $g_p g_p h_p = g_p$, 
\begin{align}
    \langle \Delta c(x, t) \Delta c(x', t') \rangle &= \frac{2 D \langle c \rangle }{L} \sum_{p=0}\sum_{p'=0}g_p g_{p'} h_p \psi_{p}(x)\psi_{p'}(x')\delta_{pp'}\nonumber \int_{-\infty}^t\int_{-\infty}^{t'}e^{-D k^2_{p}(t-s)}e^{-D k^2_{p'}(t'-s')} \delta(s-s') k^2_{p'}dsds'\nonumber\\
    &= \frac{2 D \langle c \rangle }{L} \sum_{p=0}g_p\psi_{p}(x)\psi_{p}(x')k^2_{p}\int_{-\infty}^t\int_{-\infty}^{t'}e^{-D k^2_{p}(t-s)}e^{-D k^2_{p}(t'-s')} \delta(s-s') dsds'.\nonumber
\end{align}
We now evaluate the double time integral:
\begin{align}
   \int_{-\infty}^t & \int_{-\infty}^{t'}e^{-D k^2_{ p}(t-s)}e^{-D k^2_{p}(t'-s')} \delta(s-s') dsds' =\int_{-\infty}^t e^{-D k^2_{p}(t-s)} e^{-D k^2_{p}(t'-s)}\Theta(t'-s) ds = \frac{1}{2 D k_{p}^2}e^{-D k_{p}^2|t'-t|},
\end{align}
where $\Theta$ is the Heaviside step function.  All together, we express the correlator
\begin{align}
     \langle \Delta c(x, t) \Delta c(x', t') \rangle &=\frac{\langle c \rangle }{L}\sum_{p=0}g_p\psi_{p}(x)\psi_{p}(x') e^{-D k^2_{p}|t'-t|}\nonumber\\
     &=\frac{2 \langle c \rangle }{L}\sum_{p=1}\psi_{p}(x)\psi_{p}(x') e^{-D k^2_{p}|t'-t|} \; ,
\end{align}
in agreement with Eq. \ref{eq:1d-correlator}.
\end{widetext}

\section{1D sensor calculations}
\label{app:1d-sensor-calc}
In this appendix we calculate $\delta n^2 / \langle n \rangle ^2$ 
for the six sensor geometries of interest in one dimension (Fig. \ref{fig:1d}). The expression for the sensing precision in 1D is given by Eq. \ref{eq:dn-1d} in the main text:
\begin{align}
    \frac{\delta n^2}{\langle n \rangle ^2} &= \frac{4}{\langle c \rangle L D T}\sum_{p=1}\frac{L^2}{p^2 \pi^2}\left(\frac{1}{l}\int_{\text{sensor}}\hspace{-0.6cm} \psi_p(x) dx \right)^2 .
\end{align}
Here we directly calculate $l^{-1}\int_{\text{sensor}}\psi_p(x) dx$ for each of the six sensor geometries illustrated in Fig. \ref{fig:1d} and evaluate the resulting sums. We use ``small sensor" to refer to sensors with size $\ve \rightarrow 0$, with results reported to leading order in small $\ve$, while we use ``extended sensor" to refer to sensors with finite size $a$. We also compute the correlation time $\tau$ for each sensor, which is defined according to the relationship
\begin{align}
    \frac{\delta n^2}{\langle n \rangle ^2} &= \frac{\delta n_0^2}{\langle n \rangle ^2} \frac{2\tau}{T}
\end{align}
where $\delta n_0^2 / \langle n \rangle ^2$ is the equilibrium instantaneous variance in particle number within the sensor. We directly calculate $\delta n_0^2 / \langle n \rangle ^2$ for each sensor configuration by evaluating main text Eq. \ref{eq:dn2-1d} with $t=t'$:
\begin{align}
    \frac{\delta n_0^2}{\langle n \rangle ^2}
    &= \frac{2}{\langle c \rangle L}\sum_{p=1}\left(\frac{1}{l}\int_{\text{sensor}}\hspace{-0.6cm} \psi_p(x) dx \right)^2
\end{align}
and show that the resulting variance is exactly as expected for binomial partitioning of ligand in the domain of length $L$, with probability $l/L$ of being in the sensor of total size $l$ and $1-l/L$ of being outside the sensor. 

All results from the calculations in this Appendix are summarized in main text Table \ref{tab:1d-precision}. We will use the following sum evaluations in the sections that follow:
\begin{align}
    \label{eq:l2cos2}
   &\sum_{p=1} \frac{L^2}{\pi^2 p^2}\cos^2\left(\frac{\pi a p}{L}\right) = \frac{L^2}{6}\left(1-\frac{3a}{L} + \frac{3a^2}{L^2}\right) \\
   \label{eq:l2cos2sin2}
   & \sum_{p=1} \frac{L^2}{\pi^2 p^2}\cos^2\left(\frac{\pi a p}{L}\right)\sin^2\left(\frac{\pi \ve p}{L}\right) = \frac{\ve}{4}\left(L-2\ve\right) \\
   \label{eq:l4sin2}
    &\sum_{p=1} \frac{L^4}{\pi^4 p^4}\sin^2\left(\frac{\pi a p}{L}\right) = \frac{a^2 L^2}{6}\left(1-\frac{2a}{L} + \frac{a^2}{L^2}\right)\\
    \label{eq:l2sin2}
   &  \sum_{p=1} \frac{L^2}{\pi^2 p^2}\sin^2\left(\frac{\pi a p}{L}\right) = \frac{La}{2}\left(1-\frac{a}{L}\right)\\
   \label{eq:l2cos4}
   &  \sum_{p=1} \frac{L^2}{\pi^2 p^2}\cos^4\left(\frac{\pi a p}{2 L}\right) = \frac{L^2}{6}\left(1-\frac{9a}{4L} + \frac{3a^2}{2L^2}\right)\\
   \label{eq:l2cross}
    &  \sum_{p=1}\frac{L^2}{p^2\pi^2} \left(2 \cos \left(\frac{\pi a p}{L}\right) \sin \left(\frac{\pi p \ve }{2 L}\right)+\sin \left(\frac{\pi p \ve }{L}\right)\right)^2  \nonumber\\
   & \hspace{0.5\columnwidth}=\ve\left(L-2\ve\right).
\end{align}
These sums can be checked by using trig reduction formulas for $\sin^2$ and $\cos^2$ to expand the summands, leaving expressions containing only cosine, and using the formulas \cite{G-R}:
\begin{align}
    \sum_{p=1} \frac{\cos\left(p x \right)}{p^2} &= \frac{\pi^2}{6}-\frac{\pi x}{2} + \frac{x^2}{4}\\
    \sum_{p=1} \frac{\cos\left(p x \right)}{p^4} &= \frac{\pi^4}{90}-\frac{\pi^2 x^2}{12} + \frac{\pi x^3}{12} - \frac{x^4}{48}.
\end{align}
\subsection{Centered sensors}
\subsubsection{One small sensor\label{sec:centered-small-sensor}}
We evaluate 
\begin{align}
    \frac{1}{l}\int_{\text{sensor}}\hspace{-0.6cm} \psi_p(x) dx &= \frac{1}{\ve} \int_{\frac{L-\ve}{2}}^{\frac{L+\ve}{2}}\cos\left(\frac{p \pi x}{L}\right) dx\\
    &= \frac{2L}{p\pi\ve}\cos\left(\frac{p\pi}{2}\right)\sin\left(\frac{\ve p\pi}{2 L }\right) \; .
\end{align}
The sensing precision is then
\begin{align}
     \frac{\delta n^2}{\langle n \rangle ^2} &= \frac{4}{\langle c \rangle L D T \ve^2}\sum_{p=1}\frac{4L^4}{p^4 \pi^4}\cos^2\left(\frac{p\pi}{2}\right)\sin^2\left(\frac{\ve p\pi}{2 L }\right).
\end{align}
Noting that $\cos^2(p \pi/2)$ equals 0 when $p$ is odd and 1 when $p$ is even, reindexing $p \mapsto 2p$ for $p$ even, and using Eq. \ref{eq:l4sin2},
\begin{align}
     \frac{\delta n^2}{\langle n \rangle ^2} &= \frac{1}{\langle c \rangle L D T \ve^2}\sum_{p=1}\frac{L^4}{p^4 \pi^4}\sin^2\left(\frac{\ve p\pi}{L }\right)\\
     &= \frac{L}{6\langle c \rangle  D T}\left(1-\frac{2\ve}{L} + \frac{\ve^2}{L^2}\right)\\
     &\approx \frac{L}{6\langle c \rangle  D T}, \,\,\,\,\ve \ll L .
\end{align}
The instantaneous equilibrium variance is
\begin{align}
    \frac{\delta n_0^2}{\langle n \rangle ^2} &= \frac{2}{\langle c \rangle L}\sum_{p=1}\left(\frac{1}{l}\int_{\text{sensor}}\hspace{-0.6cm} \psi_p(x) dx \right)^2\\
    &= \frac{2}{\langle c \rangle L}\sum_{p=1}\frac{1}{\ve^2}\frac{4L^2}{p^2\pi^2}\cos^2\left(\frac{p\pi}{2}\right)\sin^2\left(\frac{\ve p\pi}{2 L }\right) .
\end{align}
Reindexing $p\mapsto2p$ and using Eq. \ref{eq:l2sin2},
\begin{align}
    \frac{\delta n_0^2}{\langle n \rangle ^2} &= \frac{2}{\langle c \rangle L}\sum_{p=1}\frac{1}{\ve^2}\frac{L^2}{p^2\pi^2}\sin^2\left(\frac{\ve p\pi}{ L }\right) \\
    &= \frac{1}{\langle c \rangle \ve}\left(1-\frac{\ve}{L}\right)
\end{align}
which is the correct result for binomially partitioning $N = \avg{c} L$ molecules between the inside and outside of the sensor. In the small $\ve$ limit, the expression simplifies to
\begin{align}
    \frac{\delta n_0^2}{\langle n \rangle ^2} &\approx \frac{1}{\langle c \rangle \ve} = \frac{1}{\langle n \rangle }.
\end{align}
Therefore $\tau$ is given (in the small $\ve$ limit) by
\begin{align}
    \tau = \frac{T}{2}\frac{\langle \delta n^2\rangle / \langle n \rangle ^2}{\langle \delta n^2\rangle_0 / \langle n \rangle ^2} = \frac{L\ve}{12 D} .
\end{align}

\subsubsection{Extended sensor}
The analysis is the same as the preceding subsection, except that $\ve$ is replaced by $a$, which may have arbitrary size between $0$ to $L$. Therefore the sensing precision is
\begin{align}
     \frac{\delta n^2}{\langle n \rangle ^2} &= \frac{L}{6\langle c \rangle  D T}\left(1-\frac{2a}{L} + \frac{a^2}{L^2}\right)
\end{align}
and the equilibrium, instantaneous variance is
\begin{align}
    \frac{\delta n_0^2}{\langle n \rangle ^2} &=  \frac{1}{\langle c \rangle a}\left(1-\frac{a}{L}\right).
\end{align}
The correlation time $\tau$ is then
\begin{align}
    \tau = \frac{T}{2}\frac{\langle \delta n^2\rangle / \langle n \rangle ^2}{\langle \delta n^2\rangle_0 / \langle n \rangle ^2} = \frac{La}{12 D}\left(1-\frac{a}{L}\right).
\end{align}

\subsubsection{Two small sensors}
We evaluate 
\begin{align}
    \frac{1}{l}\int_{\text{sensor}}\hspace{-0.6cm} \psi_p(x) dx &= \frac{1}{2\ve}\bigg(\int_{\frac{L-a-\ve}{2}}^{\frac{L-a+\ve}{2}}\cos\left(k_p x\right) dx \\
    &\;\;\;\;\;\;\;\;+ \int_{\frac{L+a-\ve}{2}}^{\frac{L+a+\ve}{2}}\cos\left(k_p x\right) dx\bigg)\\
    &= \frac{2}{k_p \ve}\cos\left(\frac{a k_p}{2}\right)\cos\left(\frac{L k_p}{2}\right)\sin\left(\frac{\ve k_p}{2}\right)
\end{align}
where $k_p = p \pi / L$, and the expression for the sensing variance is, after reindexing $p\mapsto 2p$,
\begin{align}
    \frac{\delta n^2}{\langle n \rangle ^2} &= \frac{1}{\langle c \rangle L D T\ve^2}\sum_{p=1}\frac{L^4}{p^4 \pi^4}\cos^2\left(\frac{a p \pi}{L}\right)\sin^2\left(\frac{\ve p \pi}{L}\right).
\end{align}
In the limit $\ve\rightarrow0$, $L^2 \sin^2(\ve p \pi/L) / (p^2 \pi^2 \ve^2)\rightarrow 1$, and using Eq. \ref{eq:l2cos2},
\begin{align}
    \frac{\delta n^2}{\langle n \rangle ^2} &= \frac{1}{\langle c \rangle L D T}\sum_{p=1}\frac{L^2}{p^2 \pi^2}\cos^2\left(\frac{a p \pi}{L}\right)\\
     &= \frac{L}{6 \langle c \rangle  D T}\left(1-\frac{3a}{L} + \frac{3a^2}{L^2}\right).
\end{align}
The equilibrium instantaneous variance is 
\begin{align}
    \frac{\delta n_0^2}{\langle n \rangle ^2} &= \frac{2}{\langle c \rangle L}\sum_{p=1}\left(\frac{1}{l}\int_{\text{sensor}}\hspace{-0.6cm} \psi_p(x) dx \right)^2\\ 
    &= \frac{2}{\langle c \rangle L}\sum_{p=1}\frac{L^2}{p^2\pi^2 \ve^2}\cos^2\left(\frac{a p \pi}{L}\right)\sin^2\left(\frac{\ve p \pi}{L}\right)
\end{align}
after reindexing $p\mapsto2p$. Evaluating the sum (Eq. \ref{eq:l2cos2sin2}),
\begin{align}
    \frac{\delta n_0^2}{\langle n \rangle ^2} &=\frac{1}{2 \langle c \rangle \ve} \left(1-\frac{2\ve}{L}\right).
\end{align}
In the small $\ve$ limit, the correlation time is given by
\begin{align}
    \tau = \frac{T}{2}\frac{\langle \delta n^2\rangle / \langle n \rangle ^2}{\langle \delta n^2\rangle_0 / \langle n \rangle ^2} = \frac{L\ve}{6D}\left(1-\frac{3a}{L}+\frac{3a^2}{L^2}\right).
\end{align}

\subsection{Near reflecting boundary}
\subsubsection{One small sensor}
\label{sec:1d-small-sensor-boundary}
We evaluate 
\begin{align}
    \frac{1}{l}\int_{\text{sensor}}\hspace{-0.6cm} \psi_p(x) dx &= \frac{1}{\ve} \int_{\frac{a-\ve}{2}}^{\frac{a+\ve}{2}}\cos\left(\frac{p \pi x}{L}\right) dx\\
    &=  \frac{2L}{p\pi\ve}\cos\left(\frac{a p \pi}{L}\right)\sin\left(\frac{\ve p \pi}{2L}\right)
\end{align}
and the fractional variance is then
\begin{align}
    \frac{\delta n^2}{\langle n \rangle ^2} &= \frac{4}{\langle c \rangle L D T}\sum_{p=1}\frac{4L^4}{p^4 \pi^4\ve^2}\cos^2\left(\frac{a p \pi}{L}\right)\sin^2\left(\frac{\ve p \pi}{2L}\right).
\end{align}
Taking the $\ve\rightarrow 0$ limit and using Eq. \ref{eq:l2cos2},
\begin{align}
    \frac{\delta n^2}{\langle n \rangle ^2} &= \frac{4}{\langle c \rangle L D T}\sum_{p=1}\frac{L^2}{p^2 \pi^2}\cos^2\left(\frac{a p \pi}{L}\right)\\
    &= \frac{2L}{3\langle c \rangle  D T}\left(1-\frac{3a}{L} + \frac{3a^2}{L^2}\right).
\end{align}
We evaluate the instantaneous equilibrium variance using Eq. \ref{eq:l2cos2sin2}:
\begin{align}
     \frac{\delta n_0^2}{\langle n \rangle ^2}
    &= \frac{2}{\langle c \rangle L}\sum_{p=1}\left(\frac{1}{l}\int_{\text{sensor}}\hspace{-0.6cm} \psi_p(x) dx \right)^2\\
    &= \frac{2}{\langle c \rangle L}\sum_{p=1} \frac{4L^2}{p^2\pi^2\ve^2}\cos^2\left(\frac{a p \pi}{L}\right)\sin^2\left(\frac{\ve p \pi}{2L}\right)\\
     &= \frac{1}{\cavg \ve}\left(1-\frac{\ve}{L}\right)
\end{align}
and the correlation time in the small $\ve$ limit is then given by
\begin{align}
     \tau = \frac{T}{2}\frac{\langle \delta n^2\rangle / \langle n \rangle ^2}{\langle \delta n^2\rangle_0 / \langle n \rangle ^2} = \frac{L\ve}{3D}\left(1-\frac{3a}{L}+\frac{3a^2}{L^2}\right).
\end{align}

\subsubsection{Extended sensor}
We evaluate
\begin{align}
    \frac{1}{l}\int_{\text{sensor}}\hspace{-0.6cm} \psi_p(x) dx &= \frac{1}{a}\int_0^{a}\cos\left(\frac{p\pi x}{L}\right) dx\\
    &= \frac{L}{p\pi a}\sin\left(\frac{a p \pi}{L}\right)
\end{align}
and the fractional variance is then
\begin{align}
    \frac{\delta n^2}{\langle n \rangle ^2} &= \frac{4}{\langle c \rangle L D T a^2}\sum_{p=1}\frac{L^4}{p^4 \pi^4}\sin^2\left(\frac{a p \pi}{L}\right) \\
    &= \frac{2L}{3\langle c \rangle  D T}\left(1-\frac{2a}{L}+\frac{a^2}{L^2}\right)
\end{align}
using Eq. \ref{eq:l4sin2}. The equilibrium variance is
\begin{align*}
    \frac{\delta n_0^2}{\langle n \rangle ^2}
    &= \frac{2}{\langle c \rangle L}\sum_{p=1}\frac{L^2}{p^2\pi^2 a^2}\sin^2\left(\frac{a p \pi}{L}\right)\\
    &= \frac{1}{\langle c \rangle a}\left(1-\frac{a}{L}\right)
\end{align*}
using Eq. \ref{eq:l2sin2}. Therefore the correlation time is
\begin{align}
    \tau = \frac{T}{2}\frac{\langle \delta n^2\rangle / \langle n \rangle ^2}{\langle \delta n^2\rangle_0 / \langle n \rangle ^2} = \frac{La}{3D}\left(1-\frac{a}{L}\right).
\end{align}

\subsubsection{Two small sensors}
We evaluate 
\begin{align}
    \frac{1}{l}\int_{\text{sensor}}\hspace{-0.6cm} \psi_p(x) dx &= \frac{1}{2\ve}\bigg(\int_0^{\ve}\cos\left(k_p x\right) dx\nonumber \\
    &\;\;\;\;\;\;\;\;+ \int_{a-\frac{\ve}{2}}^{a+\frac{\ve}{2}}\cos\left(k_p x\right) dx\bigg)\\
    &= \frac{1}{2k_p \ve}\Bigg( \sin(k_p \ve)\nonumber\\ 
   &\;\;\;\;\;\;\;\;+ 2 \cos(k_p a)\sin\left(\frac{k_p \ve}{2}\right)\Bigg) 
\end{align}
with $k_p = p \pi / L$. In the limit $\ve\rightarrow 0$ and after applying trigonometric identities, the sensing variance is
\begin{align}
    \frac{\delta n^2}{\langle n \rangle ^2} &= \frac{4}{\langle c \rangle L D T}\sum_{p=1}\frac{L^2}{p^2 \pi^2}\cos^4\left(\frac{a p \pi}{2L}\right) \\
    &= \frac{2L}{3\langle c \rangle DT}\left(1-\frac{9a}{4L} + \frac{3a^2}{2L^2}\right)
\end{align}
using Eq. \ref{eq:l2cos4}. 
The equilibrium instantaneous variance is 
\begin{align}
    \frac{\delta n_0^2}{\langle n \rangle ^2} &= \frac{2}{\langle c \rangle L}\sum_{p=1}\left(\frac{1}{l}\int_{\text{sensor}}\hspace{-0.6cm} \psi_p(x) dx \right)^2\\ 
    &= \frac{1}{2\langle c \rangle L\ve^2} \times \\
    &\sum_{p=1}\frac{L^2}{p^2\pi^2} \left(2 \cos \left(\frac{\pi  a p}{L}\right) \sin \left(\frac{\pi  p \ve }{2 L}\right)+\sin \left(\frac{\pi  p \ve }{L}\right)\right)^2 \nonumber \\
    &= \frac{1}{2\langle c \rangle \ve} \left(1-\frac{2\ve}{L}\right)
\end{align}
using Eq. \ref{eq:l2cross}. The correlation time in the small $\ve$ limit is given by
\begin{align}
    \tau = \frac{T}{2}\frac{\langle \delta n^2\rangle / \langle n \rangle ^2}{\langle \delta n^2\rangle_0 / \langle n \rangle ^2} = \frac{2L\ve}{3D}\left(1 - \frac{9a}{4L} + \frac{3a^2}{2L^2}\right).
\end{align}

\subsection{The correlation time for small sensors\label{sec:explain-tau}}
Before turning our attention to 3D geometries, we conclude this appendix by examining in more detail the question of what sets the correlation time for small sensors.  As we already observed in the main text, all of the small sensors in Table \ref{tab:1d-precision} have correlation times that scale as $\tau \sim L \ve/D$.  An natural question is how this timescale arises from the two obvious diffusion times in the problem of order $\ve^2/D$ and $L^2/D$.  To address this issue, we recall the definition of $\tau$ (Eq. \ref{eq:tau-def}),
\begin{align}
\tau = \frac{1}{\delta n_0^2}\int_0^{\infty} \langle \Delta n(0) \Delta n(t)\rangle dt = \int_0^{\infty} G(t) dt \;,
\label{eq:tau-def2}
\end{align}
where we have defined the normalized correlation function
\begin{align}
G(t)\equiv  \frac{\avg{\Delta n(0) \Delta n(t)}}{\delta n_0^2} .
\end{align}
For illustrative purposes, we focus here on the centered small sensor, but similar arguments apply in the other small sensor cases.  Then, using the calculations from Appendix \ref{sec:centered-small-sensor}, we can write
\begin{align}
G(t) & =  \frac{1}{\delta n_0^2} \sum_{p=1} \left(\int_{\text{sensor}} \hspace{-0.6cm}\psi_p(x) dx\right)^2 e^{-D k_p^2 t} \label{eq:G-eigenfunctions} \\
& =  \frac{2 \ve}{L} \sum_{p \, \text{even}} \left(\frac{2 L}{p \pi}\right)^2 \sin^2\left(\frac{\ve p \pi}{2 L} \right) e^{-D k_p^2 t} \\
& =  \frac{2}{\barep} \sum_{p = 1} \left(\frac{1}{p \pi}\right)^2 \sin^2(p \barep \pi) e^{-4 p^2 \bart} \label{eq:define-G}\;,
\end{align}
where in the last line we have re-indexed $p \mapsto 2 p$ and introduced the dimensionless ratios $\barep = \ve/L$ and $\bart = \pi^2 D t/L^2$. (Note that, unlike in most equations earlier in Appendix \ref{app:1d-sensor-calc}, in Eq. \ref{eq:G-eigenfunctions} the integral over the sensor appears without a prefactor of $1/l$.)

We now consider how $G(t)$ behaves for different scalings of $t$. For $t$ of order $\ve^2/D$, say $\bart = \barep^2 s$ with $s \sim \mathcal{O}(1)$, we may rigorously replace the sum in Eq. \ref{eq:define-G} by an integral as $\barep \rightarrow 0$, finding
\begin{align}
G(t) = 2 \int_0^{\infty} \frac{\sin^2(\pi x)}{(\pi x)^2} e^{-4 x^2 s} dx \; \;\;\; [\bart \sim \barep^2 \;,\; \text{or} \;\; s \sim \mathcal{O}(1)].
\label{eq:small-s}
\end{align}
One can easily evaluate the integral with $s = 0$ to recover $G(t = 0) = 1$, as expected.  More generally, this expression confirms that as long as $s \sim \mathcal{O}(1)$, i.e. $t \sim \ve^2/D$, $G(t)$ is of order 1.

On the other hand, if $t$ is of order $L^2/D$, i.e. $\bart \sim \mathcal{O}(1)$, then the exponential factor causes the terms in Eq. \ref{eq:define-G} to drop off rapidly with $p$, even as $\barep \rightarrow 0$, and we may safely make the approximation $\sin^2(p \barep \pi) \approx (p \barep \pi)^2$ to obtain
\begin{align}
G(t) = 2 \barep \sum_{p = 1} e^{-4 p^2 \bart} \;\;\;\; [\bart \sim \mathcal{O}(1)] \; .
\label{eq:large-t}
\end{align}

What about intermediate values of \bart?  Conveniently, the behavior of Eq. \ref{eq:small-s} as $s \rightarrow \infty$, which can be found through a saddle point approximation, agrees with the behavior of Eq. \ref{eq:large-t} as $\bart \rightarrow 0$, which is obtained by approximating the sum by an integral.  In both cases, we find
\begin{align}
G(t) = \frac{\barep}{2} \sqrt{\frac{\pi}{\bart}} \;\;\;\; [\barep^2 \ll \bart \ll 1].
\label{eq:intermed-t}
\end{align}
A more systematic asymptotic analysis that assumes $\bart \sim \barep^\alpha$ with $0 < \alpha < 2$ arrives at the same result.  Moreover, this expression has a simple physical explanation: For times $\bart \ll 1$ (in other words $t \ll L^2/D$), diffusing particles that start at the sensor have not yet felt the effect of the reflecting boundaries a distance of order $L$ away.  Their probability of being found in the vicinity of the sensor is thus essentially governed by the usual, Gaussian Green function for diffusion in free space, which drops off like $1/\sqrt{t}$.  The prefactor of $\barep$ arises from integrating this probability over a sensor of linear size $\ve$.

Returning to the definition of $\tau$ (Eq. \ref{eq:tau-def2}),
\begin{align}
\tau = \int_0^\infty G(t) dt = \frac{L^2}{\pi^2 D} \int_0^\infty G(\bart) d\bart \; ,
\end{align}
we can evaluate the contribution of each regime to the integral over \bart.  From Eq. \ref{eq:small-s}, the integral from $\bart = 0$ to $\bart \sim \barep^2$ makes a contribution of order $\barep^2$ to $\int G(\bart) d\bart$, and thus a contribution of order $\ve^2/D$ to $\tau$.  In the other two regimes, Eqs. \ref{eq:intermed-t} and \ref{eq:large-t}, $G(t)$ can be written in the original dimensionful variables in the form $(\ve/L) f(t D/L^2)$ for some function $f$.  The integral over these regimes makes the dominant contribution to $\tau$, of order $(\ve/L) L^2/D \sim \ve L/D$.  (There is also a subdominant contribution, of order $\ve^2/D$, from the lower bound of the integral over these regimes at $t \sim \ve^2/D$.)  Thus, for small sensors, $G(t)$ has two important regimes: A power law decay, governed by the physics of free diffusion, from 1 to a value of order $\ve/L$ on timescales much less than $L^2/D$, followed by an exponential decay to zero driven by the presence of confining boundaries on timescales of order $L^2/D$. Each regime makes a contribution of order $\ve L/D$ to the correlation time $\tau$.

\section{Correlator in 3D}
\label{app:3d-correlator}
\subsection{Solution by eigenfunction expansion}
We solve the Langevin-diffusion equation for the concentration fluctuations
\begin{align}
\pdv{}{t} \Delta c(\Vec{x}, t)  &= D\nabla^2\Delta c(\Vec{x}, t)  + \eta(\Vec{x}, t)
\end{align}
with initial condition 
\begin{align}
    \Delta c(\Vec{x}, -\infty) &= 0
\end{align}
and reflecting boundary conditions in the 3 domain geometries, with 3 different symmetries, listed in Table \ref{tab:3d-geometries} and illustrated in Figs. \ref{fig:rect-cyl} and \ref{fig:spheres}. The statistics of the noise correlator are
\begin{align}
    \langle \eta(\Vec{x}, t) \eta(\Vec{x}', t') \rangle = 2 D \langle c \rangle  \delta(t-t') \nabla_{\Vec{x}}\cdot\nabla_{\Vec{x}'}\delta (\Vec{x}-\Vec{x}').
\end{align}

\begin{table*}
    \centering
    \begin{tabular}{p{0.11\textwidth} p{0.16\textwidth} p{0.41\textwidth} p{0.28\textwidth}}\toprule
    Geometry & B.C.'s & Eigenfunctions & Eigenvalues  \\
        \midrule
        Cartesian & $\displaystyle0=\pdv{\Delta c}{x_i}\Big \vert_{x_i=0,L}$  & $\displaystyle\psi_{\ell m p}=\cos\left(\frac{\ell \pi x}{L}\right)\cos\left(\frac{m \pi y}{L}\right)\cos\left(\frac{p \pi z}{L}\right)$ & $\displaystyle  k_{\ell m p}^2 = \pi^2\left(\frac{\ell^2}{L^2}+\frac{m^2}{L^2}+\frac{p^2}{L^2}\right)$\\
        \midrule
        Cylindrical &\makecell[l]{$\displaystyle 0= \pdv{\Delta c}{\rho}\Big \vert_{\rho=R}$\\ [0.4cm]$\displaystyle 0= \pdv{\Delta c}{z}\Big \vert_{z=0, L}$} & $\displaystyle \psi_{\ell m p} = J_m\left(\lambda_{m\ell}\rho\right) e^{i m \phi} \cos \left(\frac{p \pi z}{L}\right) $ &$\displaystyle k^2_{\ell m p} = \frac{p^2\pi^2}{L^2} + \lambda_{m\ell}^2$\\
        \midrule
        Spherical & $\displaystyle0=\pdv{\Delta c}{\rho}\Big \vert_{\rho=r,R}$ & \makecell[l]{$\displaystyle\psi_{\ell m p}= Y_{\ell}^m(\theta, \phi)  P_{\ell}\left(\lambda_{\ell p}\rho\right)$ \\[0.4cm]
        $\displaystyle P_{\ell}\left(\lambda_{\ell p}\rho\right) \equiv j_{\ell}\left(\lambda_{\ell p}\rho\right)+C_{\ell }y_{\ell}\left(\lambda_{\ell p}\rho\right) $ \\[0.4cm]
        $\displaystyle C_{\ell}\equiv -\frac{\ell j_{\ell}(r \lambda_{\ell p})-r \lambda_{\ell p}  j_{\ell+1}(r \lambda_{\ell p} )}{\ell y_{\ell}(r \lambda_{\ell p} )-r \lambda_{\ell p}  y_{\ell+1}(r \lambda_{\ell p} )}$}& $\displaystyle k^2_{\ell m p} =  \lambda_{\ell p}^2$\\
        \bottomrule 
        \end{tabular}
       
        \caption{\justifying To solve the Langevin-diffusion equation in three-dimensional domains with different symmetries (compare Figs. \ref{fig:rect-cyl} and \ref{fig:spheres}), we list the reflecting boundary conditions for each domain, as well as the eigenfunctions of the Laplacian adapted to the domain symmetry and their corresponding eigenvalues.  For the boundary conditions for Cartesian coordinates, $x_i = \{x,y,z\}$.}
        \label{tab:3d-geometries}
\end{table*}

As usual, we match the coordinate system to the symmetry of the domain. We write Cartesian coordinates as $(x, y, z)$, cylindrical coordinates as $(\rho, \phi, z)$, and spherical coordinates as $(\rho, \phi, \theta)$, with $\phi$ the azimuthal angle in both cylindrical and spherical coordinates and $\theta$ the polar angle. The positions of the reflecting boundaries are indicated in the second column of Table \ref{tab:3d-geometries}. Table \ref{tab:3d-geometries} also enumerates the eigenfunctions of the Laplacian, i.e. the solutions to
\begin{align}
    \nabla^2\psi &= -k^2 \psi
\end{align}
in the 3 domain geometries. The eigenfunctions are found by the standard method of separating variables \cite{Jackson1975}, and the eigenvalues by evaluating $\nabla^2 \psi_{\ell m p}$. 

In rectangular coordinates, as in 1D, $\psi_{\ell m p }$ are the cosine eigenfunctions. 

In cylindrical coordinates, the discrete spectrum $\lambda_{m \ell}$ are the values that satisfy $J'_m\left(\lambda_{m \ell}R\right)=0$, where $J_m$ are the Bessel functions of the first kind. For the radial part of the eigenfunctions, we only keep the $J_m$ solution (throwing away the $Y_m$ solution, where $Y_m$ are the Bessel functions of the second kind) because $Y_m$ diverges at $\rho=0$. 

In spherical coordinates, we keep both the $j_{\ell}$ and $y_{\ell}$ solutions, where $j_{\ell}$ and $y_{\ell}$ are respectively the spherical Bessel functions of the first and second kind, since we consider an inner reflecting boundary at finite radius $r>0$. In Table \ref{tab:3d-geometries}, we define the radial eigenfunctions in spherical coordinates $P_{\ell}\left(\lambda_{\ell p}\rho\right)$ which satisfy the no-flux boundary conditions at $\rho=r$ and $\rho=R$; in the limit $r\rightarrow 0$, it can be seen from the definition of $P_{\ell}$ that the coefficients $C_{\ell}\rightarrow 0$ so that $P_{\ell}\rightarrow j_{\ell}$. The coefficients $C_{\ell}$ are chosen to fix $P'_{\ell}(\lambda_{\ell p}r)=0$, while the discrete spectrum $\lambda_{\ell p}$ satisfy $P'_{\ell}(\lambda_{\ell p}R)=0$; i.e. they solve the transcendental equation 
\begin{align}
\label{eq:spherical-spectrum}
    &\frac{\ell j_{\ell}(\lambda_{\ell p}  r)-\lambda_{\ell p}  r j_{\ell+1}(\lambda_{\ell p}  r)}{\ell y_{\ell}(\lambda_{\ell p}  r)-\lambda_{\ell p}  r y_{\ell+1}(\lambda_{\ell p}  r)} =\\
    &\;\;\;\;\;\;\;\;\;\;\;\;\;\;\;\;\;\;\;\;\;\;\;\;\;\;\;\;\;\;\;\;\;\;\;\;\;\;\;\frac{\ell j_{\ell}(\lambda_{\ell p}  R)-\lambda_{\ell p}  R j_{\ell+1}(\lambda_{\ell p}  R)}{\ell y_{\ell}(\lambda_{\ell p}  R)-\lambda_{\ell p}  R y_{\ell+1}(\lambda_{\ell p}  R)}\nonumber.
\end{align}

We use $\lambda$ to denote the radial part of the spectrum in both spherical and cylindrical coordinates, with the choice of indices distinguishing between the two: $\lambda_{\ell p}$ in spherical domains and $\lambda_{m \ell}$ in cylindrical. 

We solve the Langevin-diffusion equation by eigenfunction expansion using the generic notation for the eigenfunctions $\psi_{\ell m p}$. The solution is to be interpreted in each particular geometry according to the definitions in Table \ref{tab:3d-geometries}. In general, the solution to the Langevin-diffusion equation is
\begin{align}
    \Delta c &= \sum_{\ell m p} a_{\ell m p}(t) \psi_{\ell m p}(\Vec{x}).
\end{align}

Substituting the expansion into the differential equation,
\begin{align}
    \sum_{\ell m p}\left( \dot{a}_{\ell m p} \psi_{\ell m p}-D\nabla^2 \psi_{\ell m p}a_{\ell m p}\right)&= \eta\nonumber\\
    \sum_{\ell m p}\left( \dot{a}_{\ell m p}+Dk_{\ell m p}^2a_{\ell m p}\right)\psi_{\ell m p}&= \eta\nonumber
\end{align}
In order to find the ODE's for the coefficients, we will make use of the orthogonality of the eigenfunctions, so we write:
\begin{align}
\label{eq:orthogonal-sum}
     \sum_{\ell m p}\left( \dot{a}_{\ell m p}+Dk_{\ell m p}^2a_{\ell m p}\right) \int_V \psi_{\ell m p}&\psi^*_{\ell' m' p'} d^3r \\
     &= \int_V \eta \psi^*_{\ell' m' p'} d^3 r.\nonumber
\end{align}
In Appendix \ref{app:3d-eig-norm} we evaluate $\int \psi_{\ell m p}{\psi^*_{\ell' m' p'}} d^3r$ explicitly in each geometry to obtain the requisite normalization coefficients. The eigenfunctions are however orthogonal for all 3 domain symmetries, so we collapse the sums in Eq. \ref{eq:orthogonal-sum} and interchange $p'\mapsto p, m'\mapsto m, \ell'\mapsto \ell$. The coefficients obey the ODEs
\begin{align}
    \dot{a}_{\ell m p}+Dk^2_{\ell m p}a_{\ell m p}&=\frac{1}{Vh_{\ell m p}}\int_V\eta(\Vec{r}, t) \psi^*_{\ell m p}(\Vec{r}) d^3 r\nonumber\\
\end{align}
where $h_{\ell m p}$ tracks the normalization of the eigenfunctions and is defined separately for each domain geometry in Appendix \ref{app:3d-eig-norm}, and $V$ is the volume of the domain, also defined in Appendix \ref{app:3d-eig-norm} for each geometry. Solving the ODEs for $a_{\ell m p}$ at steady-state (initial condition far in the past, $a_{\ell m p}(-\infty)=0$),
\begin{align}
    a_{\ell m p}(t) = \frac{1}{V h_{\ell m p}}\int_{-\infty}^t &e^{-D k^2_{\ell m p}(t-s)}\times\nonumber\\
    &\int_V \eta(\Vec{r}, s)\psi^*_{\ell m p}(\Vec{r}) d^3r\, ds.\nonumber\\
\end{align}
When $\ell=m=p=0$, $a_{000}(t)=0$, because $\psi_{000}$ is a constant (referring to Table \ref{tab:3d-geometries}, $\lambda_{00}=0$ in both spherical and cylindrical geometries) and the spatial integral over fluctuations which are conserved within the domain is $0$. Therefore we define the coefficients $g_{\ell m p}$ according to
\begin{align}
    g_{\ell m p} &= \begin{cases}
                0 & p = m = \ell = 0\\
                \dfrac{1}{h_{\ell m p}} & \text{otherwise}
                \end{cases}
\end{align}
and summarize the values of $g_{\ell m p}$ in Table \ref{tab:glmp}. With this definition of $g_{\ell m p}$ and the coefficients $a_{\ell m p}$, we express the solution for $\Delta c$:
\begin{widetext}
\begin{align}
    &\Delta c(\Vec{x},t)  =\frac{1}{V}\sum_{\ell m p}g_{\ell m p}\psi_{\ell m p}(\Vec{x})
    \int_{-\infty}^t  e^{-D k^2_{\ell m p}(t-s)}\int_V\eta(\Vec{r}, s)\psi^*_{\ell m p}(\Vec{r}) d^3r ds.
\end{align}

Now we wish to work out the correlation $\langle \Delta c(\Vec{x}, t) \Delta c(\Vec{x}', t') \rangle$. We write
\begin{align}
\label{eq:correlation}
    \langle \Delta c(\Vec{x}, t) \Delta c(\Vec{x}', t') \rangle =  \frac{1}{V^2}\sum_{\ell m p}\sum_{\ell' m' p'}&g_{\ell m p}g_{\ell' m' p'}\psi_{\ell m p}(\Vec{x})\psi^*_{\ell' m' p'}(\Vec{x'}) \int_{-\infty}^t\int_{-\infty}^{t'}e^{-D k^2_{\ell m p}(t-s)}e^{-D k^2_{\ell' m' p'}(t'-s')}\nonumber\\
    & \times \int_{V}\int_{V'}\psi^*_{\ell m p}(\Vec{r})\psi_{\ell' m' p'}(\Vec{r'})\langle \eta(\Vec{r}, s), \eta(\Vec{r}', s') \rangle d^3 rd^3 r'dsds'.
\end{align}
We will first evaluate the double spatial integral, so we substitute for the conserved noise correlator:
\begin{align}
    \int_{V}\int_{V'}\psi^*_{\ell m p}(\Vec{r})\psi_{\ell' m' p'}(\Vec{r'})\langle \eta(\Vec{r}, s), \eta(\Vec{r}', s') \rangle d^3 rd^3 r'&=2D\langle c \rangle  \delta(s-s')\int_{V}\int_{V'}\psi^*_{\ell m p}(\Vec{r})\psi_{\ell' m' p'}(\Vec{r'}) \nabla_{\Vec{r}}\cdot \nabla_{\Vec{r}'}\delta(\Vec{r}-\Vec{r}') d^3rd^3r'\nonumber\\
    &= -2D\langle c \rangle  \delta(s-s')\int_{V} \psi^*_{\ell m p}(\Vec{r}) \nabla^2\psi_{\ell' m' p'}(\Vec{r}) d^3r\nonumber
    \\
    &= 2D\langle c \rangle  \delta(s-s')k^2_{\ell' m' p'}\int_{V} \psi^*_{\ell m p}(\Vec{r}) \psi_{\ell' m' p'}(\Vec{r}) d^3r.
\end{align}
We evaluate the spatial integral as in equations \ref{eq:cartesian-volume-integral}, \ref{eq:cylindrical-volume-integral}, and \ref{eq:spherical-volume-integral}:
\begin{align}
    \int_V\psi^*_{\ell m p}(\Vec{r}) \psi_{\ell' m' p'}(\Vec{r}) d^3r &= V\delta_{\ell\ell'}\delta_{mm'}\delta_{pp'}h_{\ell m p}\nonumber.
\end{align}
Substituting into Eq. \ref{eq:correlation} and combining constant prefactors,
\begin{align}
    \langle \Delta c(\Vec{x}, t) \Delta c(\Vec{x}', t') \rangle = \frac{2 D \langle c \rangle }{V} \sum_{\ell m p}\sum_{\ell' m' p'}&g_{\ell m p}g_{\ell' m' p'}h_{\ell m p}\psi_{\ell m p}(\Vec{x})\psi^*_{\ell' m' p'}(\Vec{x'})\delta_{\ell\ell'}\delta_{mm'}\delta_{pp'}\nonumber \\
    &\times\int_{-\infty}^t\int_{-\infty}^{t'}e^{-D k^2_{\ell m p}(t-s)}e^{-D k^2_{\ell' m' p'}(t'-s')} \delta(s-s') k^2_{\ell' m' p'}dsds'\nonumber\\
    = \frac{2 D \langle c \rangle }{V} \sum_{\ell m p}g_{\ell m p}&\psi_{\ell m p}(\Vec{x})\psi^*_{\ell m p}(\Vec{x'})k^2_{\ell m p}\int_{-\infty}^t\int_{-\infty}^{t'}e^{-D k^2_{\ell m p}(t-s)}e^{-D k^2_{\ell m p}(t'-s')} \delta(s-s') dsds'\nonumber
\end{align}
where we have collapsed the double sums and used $g_{\ell m p}g_{\ell m p}h_{\ell m p}=g_{\ell m p}$ in the last line. We evaluate the double time integral as in Eq. \ref{eq:double-time-integral}:
\begin{align}
   \int_{-\infty}^t & \int_{-\infty}^{t'}e^{-D k^2_{\ell m p}(t-s)}e^{-D k^2_{\ell m p}(t'-s')} \delta(s-s') dsds' 
    =\frac{1}{2 D k_{\ell m p}^2}e^{-D k_{\ell m p}^2|t'-t|}.
\end{align}
All together, we express the correlator
\begin{align}
\label{eq:3d-correlator-app}
     \langle \Delta c(\Vec{x}, t) \Delta c(\Vec{x}', t') \rangle &=\frac{\langle c \rangle }{V}\sum_{\ell m p}g_{\ell m p}\psi^{\,}_{\ell m p}(\Vec{x})\psi^*_{\ell m p}(\Vec{x}') e^{-D k^2_{\ell m p}|t'-t|},
\end{align}
where, again, the eigenfunctions and eigenvalues $\psi_{\ell m p}$ and $k_{\ell m p}^2$ and normalization coefficients $g_{\ell m p}$ are defined separately in each domain geometry and are listed in Tables \ref{tab:3d-geometries} and \ref{tab:glmp}.

\begin{table*}
    \centering
    \begin{tabular}{c p{0.5\textwidth}}\toprule
        Geometry & Normalization coefficients $g_{\ell m p}$ \\
        \midrule
        Cartesian &  $\displaystyle\begin{cases}
                0 & p = m = \ell = 0\\
                2 & \text{two of } p, m, \ell = 0\\
                4 & \text{one of } p, m, \ell = 0\\
                8 & p, m, \ell > 0 \end{cases}$\\
        \midrule
        Cylindrical & $\displaystyle
                    \frac{{\lambda}^2_{m\ell}}{\left({\lambda}^2_{m\ell} - m^2/R^2\right)}\frac{1}{J_m^2\left(\lambda_{m\ell}R\right)} \times\begin{cases}
                0 & p = m = \ell = 0\\
                2 & p > 0\\
                1 & \text{otherwise} \end{cases}$\\
        \midrule
        Spherical & $\displaystyle \frac{(R^3-r^3)}{3\int_r^R d\rho \,\rho^2 P^2_{\ell}(\lambda_{\ell p} \rho)}
                    \times\begin{cases}
                        0 & p = m = \ell = 0\\
                        1 & \text{otherwise}
                    \end{cases}$
                    \\
        \bottomrule
    \end{tabular}
    \caption{Definition of normalization coefficients $g_{\ell m p}$ in three domain geometries.}
    \label{tab:glmp}
\end{table*}

\subsection{Eigenfunction orthogonality and normalization conditions}
\label{app:3d-eig-norm}

\subsubsection{Rectangular eigenbasis}
\label{app:orthogonality-cartesian}
We evaluate the spatial integral:
\begin{align}
\label{eq:cartesian-volume-integral}
    &\int_V\psi_{\ell m p}\psi^*_{\ell' m' p'} d^3r =\nonumber\\
    &\left(\int_{0}^{L} \cos\left(\frac{p \pi z}{L}\right)\cos\left(\frac{p' \pi z}{L}\right) dz\right)\left( \int_{0}^{L} \cos\left(\frac{m \pi y}{L}\right)\cos\left(\frac{m' \pi y}{L}\right) dy\right)\left( \int_{0}^{L} \cos\left(\frac{\ell \pi x}{L}\right)\cos\left(\frac{\ell' \pi x}{L}\right) dx\right) = \nonumber\\
    & \,\,\,\,\,\,\,\,\,\,\,\,\,\,\,\,\,\,\,\,\begin{rcases}
    \begin{dcases}
      L/2 & p = p' > 0\\
    L & p = p' = 0\\
    0 & p\neq p'
    \end{dcases}
  \end{rcases}\times
  \begin{rcases}
    \begin{dcases}
      L/2 & m = m' > 0\\
    L & m = m' = 0\\
    0 & m\neq m'
    \end{dcases}
  \end{rcases}\times
  \begin{rcases}
    \begin{dcases}
      L/2 & \ell = \ell' > 0\\
    L & \ell = \ell' = 0\\
    0 & \ell \neq \ell'
    \end{dcases}
  \end{rcases} \nonumber\\
  &\equiv V h_{\ell m p} \delta_{\ell \ell'}\delta_{mm'}\delta_{pp'}\nonumber
\end{align}
where we have defined $V\equiv L^3$ and
\begin{align}
    h_{\ell m p} = \begin{cases}
                1 & p = m = \ell = 0\\
                1/2 & \text{two of } p, m, \ell = 0\\
                1/4 & \text{one of } p, m, \ell = 0\\
                1/8 & p, m, \ell > 0.
    \end{cases}
\end{align}
to carry the factors of $\frac{1}{2}$ which result when some indices are zero.

\subsubsection{Cylindrical eigenbasis}
\label{app:orthogonality-cylindrical}
We evaluate the integral

\begin{align}
\label{eq:cylindrical-volume-integral}
\int_{V}\psi_{\ell m p}\psi^*_{\ell' m' p'} d^3r &=
      \int_{0}^{2\pi}e^{i m\phi}e^{-i m'\phi} d\phi \int_0^R \rho J_{m}\left(\lambda_{m\ell}\rho\right) J_{m'}\left(\lambda_{m'\ell'}\rho \right)  d\rho \int_{0}^{L} \cos\left(\frac{p \pi z}{L}\right)\cos\left(\frac{p' \pi z}{L}\right) dz\nonumber\\
      &= 2\pi\delta_{mm'}\delta_{\ell \ell'} \frac{\left(R^2 - m^2/\lambda_{m\ell}^2\right)}{2} J_m^2\left(\lambda_{m\ell}R\right)\times\begin{rcases}
    \begin{dcases}
      \frac{L}{2} & p = p' > 0\\
    L & p = p' = 0\\
    0 & p\neq p'
    \end{dcases}
  \end{rcases} \nonumber \\
  &\equiv V \delta_{mm'}\delta_{\ell\ell'}\delta_{pp'}h_{\ell m p}\nonumber 
\end{align}
where in the second line we have used $\delta_{m m'}$, which results from the azimuthal integral, to set $m' = m$ in the radial integral, i.e. $ J_{m'}\left(\lambda_{m'\ell'}\rho\right)\mapsto  J_{m}\left(\lambda_{m\ell'}\rho\right)$. The radial integral with reflecting boundary conditions is given in Ref. \cite{Ziener2015}. In the third line we have defined the domain volume $V=\pi R^2 L$ and the coefficient
\begin{align}
    h_{\ell m p} &= \frac{\left({\lambda}^2_{m\ell} - m^2/R^2\right)}{ {\lambda}^2_{m\ell} }J_m^2\left(\lambda_{m\ell}R\right) \times\begin{cases}
                1 & p = 0\\
                \frac{1}{2} & p > 0
    \end{cases} .\nonumber\\
\end{align}
In the main text we concern ourselves only with a cylindrically symmetric sensor, corresponding to $m=0$; for these modes the normalization simplifies to
\begin{align}
    h_{\ell m p} &= J_0^2\left(\lambda_{0\ell}R\right) \times\begin{cases}
                1 & p = 0\\
                \frac{1}{2} & p > 0
    \end{cases} .
\end{align}

\subsubsection{Spherical eigenbasis}
\label{app:orthogonality-spherical}
We evaluate the integral

\begin{align}
\label{eq:spherical-volume-integral}
\int_V\psi_{\ell m p}\psi^*_{\ell' m' p'} d^3x &=
      \int_{0}^{2\pi}\int_0^\pi Y_{\ell}^m(\theta, \phi){Y}_{\ell'}^{m'*}(\theta, \phi)\sin\theta d\theta d\phi \int_r^R \rho^2 P_{\ell}\left(\lambda_{\ell p}\rho\right)P_{\ell'}\left(\lambda_{\ell' p'}\rho\right)  d\rho  \\
      &= 4\pi\delta_{mm'}\delta_{\ell \ell'}\delta_{pp'}\int_r^R\rho^2 P_{\ell}^2\left(\lambda_{\ell p}\rho\right) d\rho\nonumber \\
      &= 4\pi\delta_{mm'}\delta_{\ell \ell'}\delta_{pp'}\left(\frac{R}{2}\left(R^2 - \frac{\ell(\ell+1)}{\lambda_{\ell p}^2}\right) P_{\ell}^2(\lambda_{\ell p}R) - \frac{r}{2}\left(r^2 - \frac{\ell(\ell+1)}{\lambda_{\ell p}^2}\right) P_{\ell}^2(\lambda_{\ell p}r)\right) \nonumber \\
  &\equiv V \delta_{mm'}\delta_{\ell\ell'}\delta_{pp'}h_{\ell m p}\nonumber
\end{align}
where we have set $\ell' = \ell$, using $\delta_{\ell \ell'}$, before performing the radial integral, i.e. $P_{\ell'}(\lambda_{\ell'p'}\rho)\mapsto P_{\ell}(\lambda_{\ell p'}\rho)$. The radial integral is given in Ref. \cite{Ziener2015}. We have defined in the last line the domain volume $V = \frac{4\pi}{3}(R^3 - r^3)$ and the normalization
\begin{align}
    h_{\ell m p} = \frac{3}{R^3-r^3}\Bigg(&\frac{R}{2}\left(R^2 - \frac{\ell(\ell+1)}{\lambda_{\ell p}^2}\right) P_{\ell}^2(\lambda_{\ell p}R) - \frac{r}{2}\left(r^2 - \frac{\ell(\ell+1)}{\lambda_{\ell p}^2}\right) P_{\ell}^2(\lambda_{\ell p}r)\Bigg)
\end{align}
In the main text we concern ourselves only with a spherically symmetric sensor, corresponding to $\ell=m=0$; for these modes the normalization simplifies to
\begin{align}
    h_{00p} &= \frac{3}{R^3-r^3}\left(\frac{R^3}{2} P_{0}^2(\lambda_{0p}R) - \frac{r^3}{2} P_{0}^2(\lambda_{0p}r)\right).
\end{align}
In the special case of the limit $r\rightarrow 0$, $h_{00p} \rightarrow \frac{3}{2} j_0^2(\lambda_{0p}R)$. By expanding $j_0$ and $j_1$ in their trigonometric definitions and setting $\sin(\lambda_{0p R})/(\lambda_{0p}R) = \cos(\lambda_{0p }R)$ to enforce the boundary condition on $j_1$, it can be shown that
\begin{align}
    j_0^2(\lambda_{0p} R) &= \frac{1}{1+\lambda_{0p}^2 R^2}. 
\end{align}
\end{widetext}
\section{3D sensor calculations}
\label{app:3d-sensor-calc}
In this appendix we calculate $\delta n^2 / \langle n \rangle ^2$ for the sensor geometries we consider in 3D (in rectangular, cylindrical, and spherical domains). The expression for the sensing precision is given by Eq. \ref{eq:dn2-3d} in the main text:
\begin{align}
    \frac{\delta n^2}{\langle n \rangle ^2} 
    &= \frac{2}{V\langle c \rangle  D T}\sum_{\ell m p}\frac{g_{\ell m p}}{k^2_{\ell m p}}\bigg|\frac{1}{v}\int_{\text{sensor}}\hspace{-0.6cm} \psi_{\ell m p}(\Vec{x}) d^3x \bigg|^2.
\end{align} 

\subsection{Quasi-1D: Cylindrical sensor}
\label{app:cyl-calc}
We define an axially symmetric sensor which extends from $\rho=0$ to $\rho=R$, as depicted in Fig. \ref{fig:3d-cyl}. The cylindrical domain has $V = \pi R^2 L$, and the sensor volume is $v = \pi R^2 a$. We evaluate
\begin{align}
    &\bigg|\frac{1}{v}\int_{\text{sensor}}\hspace{-0.6cm} \psi_{\ell m p}(\Vec{x}) d^3x \bigg|^2  \nonumber\\
    &\;\;\;\;=\frac{1}{v^2}\left|\int_0^R J_m(\lambda_{m\ell}\rho) \rho d\rho \int_{-\pi}^{\pi} e^{-i m \phi} d\phi  \int_0^a \cos\left(\frac{p\pi z}{L}\right)dz\right|^2\nonumber\\
    &\;\;\;\;= \frac{(\pi R^2)^2}{v^2} \delta_{0\ell}\delta_{0m}\dfrac{L^2}{p^2\pi^2}\sin^2\left(\dfrac{a p \pi}{L}\right).
\end{align}
The sensing precision is thus
\begin{align}
    \frac{\delta n^2}{\langle n \rangle ^2} 
    &= \frac{2}{V\langle c \rangle  D T}\frac{(\pi R^2)^2}{v^2} \sum_{p=1}\frac{g_{00p}}{k^2_{00p}}\frac{L^2}{p^2\pi^2}\sin^2\left(\dfrac{a p \pi}{L}\right) \\
    &= \frac{4}{LDT\langle c \rangle \pi R^2 a^2} \sum_{p=1}\frac{L^4}{p^4\pi^4}\sin^2\left(\dfrac{a p \pi}{L}\right)
\end{align} 
using $g_{00p} = 2/J_0^{2}(0) = 2$ and $k_{00p}^2 = p^2\pi^2 / L^2$. This is the same sum as Eq. \ref{eq:l4sin2} which arose in 1D geometries, so we have turned our 3D problem into a quasi-1D problem.  This is not surprising, because whether molecules are found within the sensor depends only only their diffusion along the $z$ axis, which is independent from their diffusion along orthogonal directions. Evaluating the sum,
\begin{align}
   \frac{\delta n^2}{\langle n \rangle ^2}   &= \frac{2L}{3DT(\pi R^2\langle c \rangle )}\left(1-\frac{2a}{L} + \frac{a^2}{L^2}\right)
\end{align}
which is the same result as for the extended sensor near a reflecting boundary in 1D with $\langle c \rangle$ replaced by $\pi R^2 \langle c \rangle $. 

\subsection{Small cubic sensors}
\label{app:small-rect-calc}
We consider a cubic domain with edge length $L$ and a cubic sensor of edge length $a$, which we place at four positions in the domain, illustrated in Fig. \ref{fig:3d-small}. In contrast to other geometries where we have allowed $a$ to take on arbitrary values, here we only consider small sensors with $a\ll L$. This subsection has two purposes. First, as in the other subsections in this appendix, we wish to calculate the sensing precision $\delta n^2/\avg{n}^2$ for these four sensors. Second, in order to make contact with standard results for a small sensor in an infinite, three-dimensional domain \cite{Berg1977, Kaizu2014, Aquino2015}, we would like to be able to consider the limit $L \rightarrow \infty$ while $a$ and the averaging time $T$ remain finite. Our development in the main text, however, assumed that $T \gg L^2/D$ (see, e.g., the text below Eq. \ref{eq:time-avg}).  Here, we will show that for cubic sensors in three dimensions it is sufficient to require $T \gg a^2/D$ to reach the long averaging time limit where $\delta n^2 \sim T^{-1}$, and we will discuss why the more stringent condition $T \gg L^2/D$ is needed in one dimension.

\subsubsection{Long-time limit for cubic sensors in 3D}
For the four rectangular sensors illustrated in Fig. \ref{fig:3d-small}, the limits of integration are
\begin{enumerate}
    \item $x\in [\frac{L-a}{2}, \frac{L+a}{2}], y\in [\frac{L-a}{2}, \frac{L+a}{2}], z\in [\frac{L-a}{2}, \frac{L+a}{2}]$
    \item $x\in [\frac{L-a}{2}, \frac{L+a}{2}], y\in [\frac{L-a}{2}, \frac{L+a}{2}], z\in [0, a]$
    \item $x\in [\frac{L-a}{2}, \frac{L+a}{2}], y\in [0, a], z\in [0, a]$
    \item $x\in [0, a], y\in [0, a], z\in [0, a]$.
\end{enumerate}
We define $I^2_{\ell m p}\equiv \big|\frac{1}{v}\int_{\text{sensor}}\psi_{\ell m p}(\Vec{x}) d^3x \big|^2  \nonumber$ and evaluate the integral for each sensor:
\begin{align}
    {I_{\ell m p}^{2 (1)}} = \frac{64L^6}{\pi^6 \ell^2 m^2 p^2 a^6}&\cos^2\left(\frac{\ell \pi}{2}\right)\cos^2\left(\frac{m \pi}{2}\right)\cos^2\left(\frac{p \pi}{2}\right)\times\nonumber  \\
    & \sin^2\left(\frac{\ell \pi a}{2L}\right)\sin^2\left(\frac{m \pi a}{2L}\right)\sin^2\left(\frac{p \pi a}{2L}\right) \nonumber,\\
    {I_{\ell m p}^{2 (2)}} = \frac{16L^6}{\pi^6 \ell^2 m^2 p^2 a^6}&\cos^2\left(\frac{\ell \pi}{2}\right)\cos^2\left(\frac{m \pi}{2}\right)\times\nonumber  \\
    & \sin^2\left(\frac{\ell \pi a}{2L}\right)\sin^2\left(\frac{m \pi a}{2L}\right)\sin^2\left(\frac{p \pi a}{L}\right) \nonumber,\\
    {I_{\ell m p}^{2 (3)}} = \frac{4L^6}{\pi^6 \ell^2 m^2 p^2 a^6}&\cos^2\left(\frac{\ell \pi}{2}\right)\times\nonumber  \\
    & \sin^2\left(\frac{\ell \pi a}{2L}\right)\sin^2\left(\frac{m \pi a}{L}\right)\sin^2\left(\frac{p \pi a}{L}\right) \nonumber,\\
    {I_{\ell m p}^{2 (4)}} = \frac{L^6}{\pi^6 \ell^2 m^2 p^2 a^6}&\sin^2\left(\frac{\ell \pi a}{L}\right)\sin^2\left(\frac{m \pi a}{L}\right)\sin^2\left(\frac{p \pi a}{L}\right) \nonumber.
\end{align}
Defining 
\begin{align}
     {\sigma_{\ell m p}^{2}} &\equiv \frac{L^6}{\pi^6 \ell^2 m^2 p^2a^6}\sin^2\left(\frac{\ell \pi a}{L}\right)\sin^2\left(\frac{m \pi a}{L}\right)\sin^2\left(\frac{p \pi a}{L}\right)
\end{align}
and reindexing using the $\cos^2$ factors for the four sensors, we use main text Eq. \ref{eq:dnt-3d} to write the time correlators for number occupancy in the four sensors:
\begin{align}
    \frac{\langle \Delta n(t) \Delta n(t')  \rangle_1 }{\langle n \rangle ^2} &=  \frac{1}{V\langle c \rangle }\sum_{\ell m p}g_{\ell m p}{\sigma_{\ell m p}^{2 }}e^{-Dk^2_{2\ell, 2 m, 2p}|t-t'|}\nonumber\\
    \frac{\langle \Delta n(t) \Delta n(t')  \rangle_2 }{\langle n \rangle ^2} &=  \frac{1}{V\langle c \rangle }\sum_{\ell m p}g_{\ell m p}{\sigma_{\ell m p}^{2 }}e^{-Dk^2_{2\ell, 2 m, p}|t-t'|}\nonumber\\
     \frac{\langle \Delta n(t) \Delta n(t')  \rangle_3 }{\langle n \rangle ^2} &=  \frac{1}{V\langle c \rangle }\sum_{\ell m p}g_{\ell m p}{\sigma_{\ell m p}^{2 }}e^{-Dk^2_{2\ell, m, p}|t-t'|}\nonumber\\
      \frac{\langle \Delta n(t) \Delta n(t')  \rangle_4 }{\langle n \rangle ^2} &=  \frac{1}{V\langle c \rangle }\sum_{\ell m p}g_{\ell m p}{\sigma_{\ell m p}^{2 }}e^{-Dk^2_{\ell m p}|t-t'|}\nonumber.
\end{align}
To evaluate the sensing precision, we need to evaluate the time average $T^{-2}\int_0^T\int_0^T \langle \Delta n(t) \Delta n(t')  \rangle dt dt'$. We will explicitly show the calculation for sensor 4, but the procedure is identical for the other sensors, with the appropriately re-indexed eigenvalues. Directly evaluating the time integral,

\begin{align}
    \frac{\delta n^2_4 }{\langle n \rangle ^2} =\frac{1}{V\langle c \rangle }\sum_{\ell m p}g_{\ell m p}&{\sigma_{\ell m p}^{2 }}\times \\
    &\left(\frac{2}{D k^2_{\ell m p} T} +\frac{2(e^{-Dk^2_{\ell m p}T}-1)}{D^2 k_{\ell m p}^4 T^2}\right)\nonumber.
\end{align}
How large does $T$ need to be for this sum to approach its asymptotic, large $T$ behavior? Clearly, $T \gg 1/(Dk_{111}^2)$ is sufficient (as we argued for one-dimensional systems in the main text). However, when the sensor is small, we will show that averaging times this long are not necessary; all that is necessary in three dimensions is for the integration time to be longer than the correlation time set by the sensor size, $T\gg a^2 / D$. 

We can see this by approximating the sum by an integral in the limit $a\ll L$. We define the dimensionless $q_x = \ell \pi a/L$ (and likewise for $m$ and $p$ in respectively the $y$ and $z$ directions), with small $\Delta q = a\pi/L$. With these definitions for $q_x$, $q_y$, and $q_z$, $\sigma^2_{\ell m p} = \sin^2\left(q_x\right)\sin^2\left(q_y\right)\sin^2\left(q_z\right) / (q_x^2 q_y^2 q_z^2)$. Then we rewrite the sums and take the $\Delta q\rightarrow 0$ limit. The sensing precision in this limit is
\begin{align}
\label{eq:variance-integral-3d-full}
    \frac{\delta n^2_4 }{\langle n \rangle ^2} = \frac{8}{a^3 \pi^3 \cavg} &\int d^3 q \frac{\sin^2(q_x)\sin^2(q_y)\sin^2(q_z)}{q_x^2 q_y^2 q_z^2} \\
    &\times \left(\frac{2}{D |q|^2 T / a^2} +\frac{2(e^{-D |q|^2 T/a^2}-1)}{D^2 |q|^4 T^2/a^4}\right) \nonumber
\end{align}
where the factor $8$ comes from $g_{\ell m p}$ when each of $\ell, m, p >1$, and the integral is taken only over the positive octant. Using the definition of $g_{\ell m p}$ for the rectangular coordinate system (Table \ref{tab:glmp}), the contributions to the triple sum $\sum_{\ell m p}$ can be broken into contributions for each one of the indices 0, each pair of indices 0, and all indices nonzero. Only the contributions for all indices nonzero (for which $g_{\ell m p }=8$) contribute in the small $a$ limit. In this limit, the contributions when one index is zero go like $L^{-1}$, the contributions when two indices are zero go like $a L^{-2}$, and the contributions for each index $\geq 1$ go like $a^{-1}$. Therefore the latter contributions dominate when $a \ll L$.  

Taking the limit $T\gg a^2/D$ in Eq. \ref{eq:variance-integral-3d-full} and implicitly assuming that the integral is dominated by $q$ of order 1, we can simplify the expression for the sensing variance to
\begin{align}
\label{eq:variance-integral-3d-limit}
    \frac{\delta n^2_4 }{\langle n \rangle ^2} = \frac{16}{\pi^3 a\langle c \rangle DT} \int  \frac{\sin^2(q_x)\sin^2(q_y)\sin^2(q_z)}{(q_x^2 + q_y^2 + q_z^2) q_x^2  q_y^2  q_z^2} d^3q
\end{align}
which converges on $q_i \in \{0, \infty\}$ for $i = \{x, y, z\}$.  This convergence is \textit{a posteriori} justification that values of $q$ of order 1, rather than very small $q$, dominate the integral and thus that the asymptotic large $T$ behavior is obtained when $T \gg a^2/D$.

Calculating the expressions for the other sensors,
\begin{align}
    \frac{\delta n^2_1}{\langle n \rangle ^2}
    &=  \frac{16}{\pi^3 a\langle c \rangle DT} \int  \frac{\sin^2(q_x)\sin^2(q_y)\sin^2(q_z)}{4(q_x^2 + q_y^2 + q_z^2) q_x^2  q_y^2  q_z^2} d^3q \nonumber\\
     \frac{\delta n^2_2}{\langle n \rangle ^2}
     &=  \frac{16}{\pi^3 a\langle c \rangle DT} \int  \frac{\sin^2(q_x)\sin^2(q_y)\sin^2(q_z)}{(4q_x^2 + 4q_y^2 + q_z^2) q_x^2  q_y^2  q_z^2} d^3q \nonumber\\
     \frac{\delta n^2_3}{\langle n \rangle ^2}
     &=  \frac{16}{\pi^3 a\langle c \rangle DT} \int  \frac{\sin^2(q_x)\sin^2(q_y)\sin^2(q_z)}{(4q_x^2 + q_y^2 + q_z^2) q_x^2  q_y^2  q_z^2} d^3q \nonumber
\end{align}
which differ only by the denominators in the integrals. The integrals contribute only constants; the prefactor for the sensing precision increases with sensor index (increases with number of reflecting boundaries), with $\delta n_4^2 /\langle n \rangle ^2$ the largest, $\delta n_1^2 /\langle n \rangle ^2$ the smallest and $\delta n_4^2 /\langle n \rangle ^2 = 4\delta n_1^2 /\langle n \rangle ^2$. We numerically evaluate the integrals and report the prefactors in Table \ref{tab:3d-cubes}.

\subsubsection{Comparison to long-time limit in 1D}
Here, we show how the approach in the preceding section fails in 1D, so that we must require $T\gg L^2/D$ for $\delta n^2 \sim T^{-1}$ in 1D. Using the results from Appendix \ref{sec:1d-small-sensor-boundary}, we write the expression for the time correlation function for the centered small sensor of size $\ve$ (similar arguments will hold for other small sensors):
\begin{align}
    \frac{\langle \Delta n(t) \Delta n(t')\rangle }{\langle n \rangle^2} = \frac{2}{L\cavg }\sum_{p=1} & \frac{\sin^2\left(\ve k_p\right)}{k_p^2 \ve^2} e^{-4Dk_p^2|t-t'|}
\end{align}
where the eigenvalues $k_p^2 = p^2 \pi^2/L^2$. Taking the time average,
\begin{align}
    \frac{\delta n^2}{\langle n \rangle^2} = \frac{2}{L\cavg }\sum_{p=1}& \frac{\sin^2\left( k_p\ve \right)}{k_p^2 \ve^2} \left(\frac{2}{D k_p^2 T} +\frac{2(e^{-Dk^2_p T}-1)}{D^2 k_p^4 T^2}\right).
\end{align}
It is clear that $T\gg L^2/D$ is sufficient for $\delta n^2 \sim T^{-1}$ in the long-time limit. Why is $T\gg \ve^2/D$ not sufficient? We will show this by approximating the sum by an integral for small $\ve \ll L$, as we did in 3D. We define $q \equiv \ve k_p$ and $\Delta q \equiv \pi \ve/L$. Taking the $\Delta q\rightarrow 0$ limit,
\begin{align}
\label{eq:1d-int-small-sensor}
    \frac{\delta n^2}{\langle n \rangle^2} = \frac{2}{\cavg \ve}\int_0^{\infty} & \frac{dq}{q^2}\sin^2(q) \\
    \times &\left(\frac{2}{D q^2 T / \ve^2} +\frac{2(e^{-D q^2 T/\ve^2}-1)}{D^2 q^4 T^2/\ve^4}\right)\nonumber.
\end{align}
This is expression is finite as $q \rightarrow 0$. If however we attempt to naively take $T\gg \ve^2/D$ with $q$ of order 1, as we did in 3D, the resulting integral reads
\begin{align}
    \frac{\delta n^2}{\langle n \rangle^2} = \frac{4 \ve }{\cavg D T }\int_0^{\infty} & \frac{dq}{q^4}\sin^2(q) .
\end{align}
Unlike for the corresponding 3D limit (Eq. \ref{eq:variance-integral-3d-limit}), this expression diverges at small $q$. This is the same divergence pointed out in Ref. \cite{Tkacik2009} in 1D infinite space. In 1D, unlike in 3D, the dominant contributions to $\delta n^2$ are the long-wavelength contributions, and thus long averaging times $T\gg L^2/D$ are required in order for $\delta n^2 \sim T^{-1}$. The integral in Eq. \ref{eq:1d-int-small-sensor} can be directly evaluated symbolically for arbitrary $T D / \ve^2$ using Mathematica \cite{Mathematica}, and when $T\gg \ve^2 / D$ the sensing precision has the limiting form
\begin{align}
    \frac{\delta n^2}{\langle n \rangle^2} &= \frac{8\sqrt{\pi}}{3}\frac{1}{\cavg \sqrt{DT}},
\end{align}
showing the expected $T^{-1/2}$ scaling for a sensor in free space \cite{Bicknell2015, Tkacik2009}, which we expect to hold until $T \sim L^2/D$.

\subsection{Spherical sensors}
\label{app:sphere-calcs}
In this section, we focus on the sensor geometries with spherical symmetry shown in Fig. \ref{fig:sphere-models}.  Unlike for the cubic sensors of the preceding section, here for simplicity we do not directly address the question of what timescale $T$ must be compared to in order to determine whether $T$ is large. Based on the intuition developed from the cubic sensors, however, we expect that when $R \gg a$ it will be enough to have $T \gg a^2/D$.

\subsubsection{Extended sensor (ES)}
The extended sensor extends from $\rho=0$ to $\rho=a$ and is spherically symmetric, with volume $v=4\pi a^3/3$ and domain volume $V=4\pi R^3/3$. We evaluate the volume integral
\begin{align}
    &\bigg|\frac{1}{v}\int_{\text{sensor}}\hspace{-0.6cm} \psi_{\ell m p}(\Vec{x}) d^3x \bigg|^2 \nonumber \\
    &=\frac{1}{v^2}\left|\int_{0}^{2\pi} d\phi \int_{0}^{\pi}  Y_{\ell}^m(\theta, \phi)\sin\theta d\theta \int_{0}^{a} j_{\ell}(\lambda_{\ell p} \rho) \rho^2 d\rho\right|^2\nonumber \\
    &= \frac{(4\pi)^2 \delta_{0\ell}\delta_{0m}}{v^2} \left(\int_{0}^{a} j_{0}(\lambda_{0  n} \rho) \rho^2 d\rho\right)^2\nonumber\\
    &= 9 \delta_{0\ell}\delta_{0m} \frac{j_1^2 (\lambda_{0p} a)}{\lambda_{0p}^2 a^2}
\end{align}
where the third line follows from the fact that  we defined the normalization condition of the $Y_{\ell}^m$ such that $Y_0^0=1$ (Eq. \ref{eq:spherical-volume-integral}), and we then evaluate the solid angle integral by using the orthogonality condition of the $Y_{\ell}^m$. 

The sensing precision is given by
\begin{align}
     \frac{\delta n^2}{\langle n \rangle ^2} 
    &= \frac{18}{V\langle c \rangle  D T}\sum_{p=1}\frac{g_{00p}}{k^2_{00p}}   \frac{j_1^2 (\lambda_{0p} a)}{\lambda_{0p}^2 a^2}\\
    &= \frac{12 R^4}{V \langle c \rangle  DTa^2} \sum_{p=1} \frac{j_1^2 (z_p a/R)}{j_0^2(z_p)} \frac{1}{z_p^4 }
\end{align}
after using $g_{00p} = 2/(3j_0^2(\lambda_{0p}R))$, $k^2_{00p} = \lambda_{0p}^2$, and defining $z_p \equiv \lambda_{0p}R$.  To evaluate the sum
\begin{align}
    g(x) \equiv \sum_{p=1} \frac{j_1^2(z_p x)}{j_0^2(z_p)}\frac{1}{z_p^4} 
\end{align}
(with $x\equiv a/R$) exactly, we will combine Feynman's trick (differentiating with respect to the parameter $x$ under the sum operator) with the approach based on Fourier-Bessel series introduced in Ref. \cite{Bicknell2015}. First, differentiating $j_1^2(x z_p)$ with respect to $x$, 
\begin{align}
    \pdv[]{}{x} j_1^2(x z_p) &= -\frac{4}{x} j_1^2(x z_p) + 2 z_p j_0(x z_p)j_1(x z_p)\;\;\;\;\;\;\;\;\;\;\;\;
\end{align}
and therefore
\begin{align}
    g'(x) &= -\frac{4}{x}g(x) + \sum_{p=1} \frac{2 j_0(x z_p)j_1(x z_p)}{j_0^2(z_p) z_p^3}.
\end{align}
Next, we realize that 
\begin{align}
    2 z_p j_0(x z_p)j_1(x z_p) &= -\pdv{}{x} j_0^2(x z_p)
\end{align}
and we rewrite the differential equation
\begin{align}
     g'(x) &= -\frac{4}{x}g(x) -\pdv[]{}{x} \sum_{p=1} \frac{j_0^2(x z_p)}{j_0^2(z_p) z_p^4}.
\end{align}
The sum is similar to that studied in Ref. \cite{Bicknell2015}, except the terms decrease with $z_p^{-4}$ rather than $z_p^{-2}$. To evaluate this new sum, we will use the same procedure based on Fourier-Bessel series, with minor modification. We first use trig reduction to expand the spherical Bessel function:
\begin{align}
    \sum_{p=1} \frac{j_0^2(x z_p)}{j_0^2(z_p) z_p^4} &= \frac{1}{2x^2}\sum_{p=1} \frac{1}{j_0^2(z_p) z_p^6}\left(1 - \cos(2 x z_p)\right)\nonumber\\
    &= \frac{1}{2x^2}\sum_{p=1} \frac{1+z_p^2}{ z_p^6}-  \frac{1}{2x^2}\sum_{p=1} \frac{\cos(2 x z_p)}{j_0^2(z_p) z_p^6} .\nonumber\\
\end{align}
To evaluate the sum
\begin{align}
    f(x) \equiv \sum_{p=1} \frac{\cos(x z_p)}{j_0^2(z_p) z_p^6},
\end{align}
we follow the approach of Ref. \cite{Bicknell2015} and consider the related sum 
\begin{align}
    h(x) \equiv \sum_{p=1} \frac{j_1(x z_p)}{j_0^2(z_p) z_p^7},
\end{align}
which has the property that
\begin{align}
    f(x) = \pdv{}{x}\left(\frac{1}{x} \pdv{x}\left(x^2 h(x) \right)\right).
\end{align}
To evaluate $h(x)$, we realize that the sum represents a Fourier-Bessel series in $j_1$; we use the orthogonality of $j_1$ on $[0, 1]$ to write
\begin{align}
    \int_0^1 h(x) &j_1 (x z_{p'}) x^2 dx\nonumber\\
    &= \sum_{p=1} \frac{1}{j_0^2(z_p) z_p^7} \int_0^1 j_1(x z_p) j_1 (x z_{p'}) x^2 dx\nonumber\\
    &= \sum_{p=1} \frac{1}{j_0^2(z_p) z_p^7} \frac{1}{2(1+z_p^2)}\delta_{pp'}\nonumber\\
    &= \frac{1}{2 z_{p'}^7}.
\end{align}
Using a polynomial ansatz for $h(x)$ and solving for the coefficients by equating at each order in $z_p$, we determine that $h(x)$ is given by 
\begin{align}
    h(x) &= \frac{x^7}{30240}-\frac{3 x^5}{2800}+\frac{x^4}{288}-\frac{3 x^3}{875}+\frac{47 x}{47250}\;\;\;\;\;\;\;\;
\end{align}
and therefore
\begin{align}
    f(x) &= \frac{x^6}{480}-\frac{3 x^4}{80}+\frac{x^3}{12}-\frac{9 x^2}{175}+\frac{47}{15750}.
\end{align}
Therefore returning to our sum
\begin{align}
    \sum_{p=1} \frac{j_0^2(x z_p)}{j_0^2(z_p) z_p^4} 
    &= \frac{1}{2x^2}\sum_{p=1} \frac{1+z_p^2}{ z_p^6}-  \frac{1}{2x^2}f(2x)\\
    &=  \frac{1}{2x^2}\sum_{p=1} \left(\frac{1}{z_p^6} + \frac{1}{z_p^4}\right)-  \frac{1}{2x^2}f(2x).\nonumber
\end{align}
Using the formulas for the sum over zeros of $j_1$ (zeros of $J_{3/2}$) \cite{Elizalde1993},
\begin{align}
    \sum_{p=1}\frac{1}{z_p^6} &= \frac{1}{32(\frac{3}{2}+1)^3(\frac{3}{2}+2)(\frac{3}{2}+3)} = \frac{1}{7875}\\
     \sum_{p=1}\frac{1}{z_p^4} &= \frac{1}{16(\frac{3}{2}+1)^2(\frac{3}{2}+2)} = \frac{1}{350}.
\end{align}
Then we evaluate $f(2x) / (2x^2)$ and write for our sum:
\begin{align}
     \sum_{p=1} \frac{j_0^2(x z_p)}{j_0^2(z_p) z_p^4} 
    &= -\frac{x^4}{15}+\frac{3 x^2}{10}-\frac{x}{3}+\frac{18}{175}.
\end{align}
Therefore the differential equation for the original sum, $g(x)$, is
\begin{align}
     g'(x) &= -\frac{4}{x}g(x) -\pdv[]{}{x} \left(-\frac{x^4}{15}+\frac{3 x^2}{10}-\frac{x}{3}+\frac{18}{175}\right)\nonumber \\
     &= -\frac{4}{x}g(x) + \frac{4 x^3}{15}-\frac{3 x}{5}+\frac{1}{3}
\end{align}
Together with the boundary condition $g(1)=0$ (which is apparent from the definition of $g(x)$ since $j_1(z_p)=0$), the solution for $g(x)$ is
\begin{align}
    g(x) = \frac{x^4}{30}-\frac{x^2}{10}+\frac{x}{15}.
\end{align}
Therefore the expression for the fractional variance is
\begin{align}
    \frac{ \delta n^2}{\langle n \rangle ^2} &= \frac{3}{5 \pi  a \langle c \rangle  D T} \left(1-\frac{3 a}{2 R}+\frac{a^3}{2 R^3}\right).
\end{align}
In the limit $a\rightarrow 0$, this is the Berg-Purcell result for the extended sensor. In the limit $a\rightarrow R$, the fractional variance goes to zero, as it must when all molecules are enclosed by the sensor.

\subsubsection{Permeable shell (PS)}
In this model, $r=0$, and we consider a thin sensing shell that extends from $\rho = a$ to $a+\ve$. The sensor volume is  $v = \frac{4\pi}{3} ((a+\ve)^3 - a^3)$ with $\ve \ll a$, and the domain volume is $V=4\pi R^3/3$.

We evaluate the volume integral
\begin{align}
    &\bigg|\frac{1}{v}\int_{\text{sensor}}\hspace{-0.6cm} \psi_{\ell m p}(\Vec{x}) d^3x \bigg|^2 \nonumber \\
    &=\frac{1}{v^2}\left|\int_{0}^{2\pi} d\phi \int_{0}^{\pi}  Y_{\ell}^m(\theta, \phi)\sin\theta d\theta \int_{a}^{a+\ve} j_{\ell}(\lambda_{\ell p} \rho) \rho^2 d\rho\right|^2\nonumber \\
    &= \frac{(4\pi)^2 \delta_{0\ell}\delta_{0m}}{v^2} \left(\int_{a}^{a+\ve} j_{\ell}(\lambda_{\ell  n} \rho) \rho^2 d\rho\right)^2\nonumber.
\end{align}
Since $\delta_{0\ell}$ enforces $\ell=0$ we evaluate the limit
\begin{align}
    \lim_{\ve\rightarrow 0} \frac{1}{v^2}\left(\int_{a}^{a+\ve} j_{0}(\lambda_{0p} \rho) \rho^2 d\rho\right)^2 = \left(\frac{j_0\left(\lambda_{0p} a\right)}{4 \pi}\right)^2,
\end{align}
and we write the volume integral
\begin{align}
    \bigg|\frac{1}{v}\int_{\text{sensor}}\hspace{-0.6cm} \psi_{\ell m p}(\Vec{x}) d^3x \bigg|^2 &= \delta_{0\ell}\delta_{0m}  j_0^2\left(\lambda_{0p} a\right).
\end{align}
Then the expression for the fractional variance is
\begin{align}
    \frac{\delta n^2}{\langle n \rangle ^2} 
    &= \frac{2}{V\langle c \rangle  D T}\sum_{p=1}\frac{g_{00p}}{k^2_{00p}}j_0^2\left(\lambda_{0p} a\right)\\
    \label{eq:PS-sum}
    &= \frac{1}{\pi R \langle c \rangle DT}\sum_{p=1}\frac{1}{z_p^2}\frac{j_0^2\left(z_p a/R\right)}{j_0^2\left(z_p\right)}
\end{align} 
after using $g_{00p} = 2/(3j_0^2(\lambda_{0p}R))$, $k^2_{00p} = \lambda_{0p}^2$, and defining $z_p \equiv \lambda_{0p}R$. This is the same sum that appears in Ref. \cite{Bicknell2015}, in which the eigenfunction normalization is notated $||j_0^2|| \equiv \frac{1}{2(1+z_p^2)} = j_0^2(z_p)/2$. Using this notation we write the sum
\begin{align}
    \frac{\delta n^2 }{\langle n \rangle ^2} &=\frac{1}{ 2 \pi R  \langle c \rangle  DT} \sum_{p=1} \frac{j_0^2\left(z_p a/ R\right) }{|| j_0^2||z_p^2} 
\end{align}
which allows us to read off the solution for the sum from Ref. \cite{Bicknell2015}:
\begin{align}
\label{eq:PS-solution}
    \frac{\delta n^2 }{\langle n \rangle ^2} &=\frac{1}{ 2 \pi a  \langle c \rangle  DT} \left(1 - \frac{9a}{5 R} + \frac{a^3}{R^3}\right).
\end{align}
Thus the PS model corresponds exactly to the diffusion floor for a shell of receptors obtained in Ref. \cite{Bicknell2015} (for arbitrary $a/R$) and Ref. \cite{Berezhkovskii2013} (for small $a/R$). 

\subsubsection{Impermeable shell (IS)}
\begin{figure*}
    \centering
\includegraphics{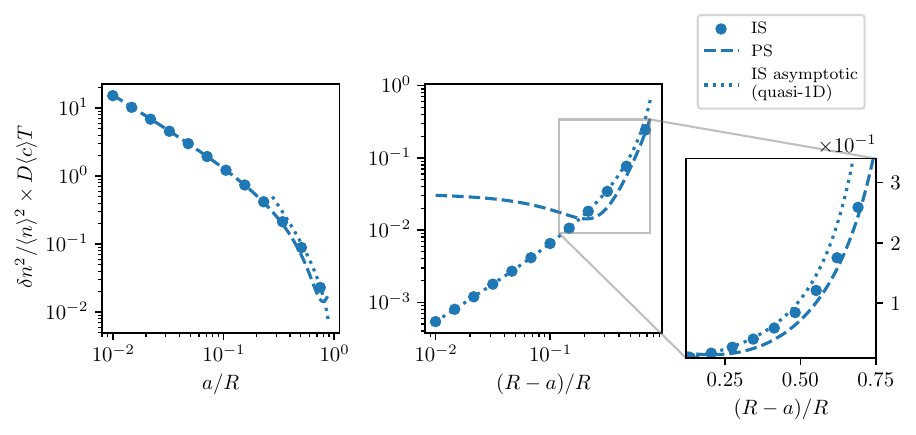}
    \caption{\justifying IS sensing precision. Large dots, exact solution (evaluated numerically); dashed line, exact expression for the PS; dotted asymptotic expression for the IS in the quasi-1D, small $R-a$ limit (Eq. \ref{eq:IS-large}).  The IS solution asymptotically approaches the PS solution in the small $a$ limit. In the small $R-a$ limit, the exact solution asymptotically approaches the quasi-1D expression. Inset: Sensing precision for intermediate $a/R$. }
    \label{fig:IS-interp}
\end{figure*}

In this model, $r>0$, and we consider a thin sensing shell which extends from $\rho=r$ to $\rho = a$ with $a-r = \ve \ll a$. The volume of the sensing shell is $v = \frac{4\pi}{3} \left(a^3 - r^3\right)$ and the domain volume is $V = \frac{4\pi}{3} \left(R^3 - r^3\right)$. We evaluate the volume integral
\begin{align}
    &\bigg|\frac{1}{v}\int_{\text{sensor}}\hspace{-0.6cm} \psi_{\ell m p}(\Vec{x}) d^3x \bigg|^2 \nonumber \\
    &=\frac{1}{v^2}\left|\int_{0}^{2\pi} d\phi \int_{0}^{\pi}  Y_{\ell}^m(\theta, \phi)\sin\theta d\theta \int_{r}^{a} P_{\ell}(\lambda_{\ell p} \rho) \rho^2 d\rho\right|^2\nonumber \\ 
    &=\frac{(4\pi)^2 \delta_{0\ell}\delta_{0m}}{v^2}\left(\int_{r}^{a} P_{\ell}(\lambda_{\ell p} \rho) \rho^2 d\rho\right)^2.
\end{align}
Since $\delta_{0\ell}$ enforces $\ell=0$, we evaluate the integral
\begin{align}
   \int_{r}^{a}& P_{0}(\lambda_{0  n} \rho) \rho^2 d\rho \\
    &= \int_{r}^{a} \left(j_0(\lambda_{0p} \rho)-\frac{j_1(\lambda_{0p} r)}{y_1(\lambda_{0p} r)}y_0(\lambda_{0p} \rho)\right) \rho^2 d\rho\nonumber\\
    &=  \left(j_1(\lambda_{0p} \rho)-\frac{j_1(\lambda_{0p} r)}{y_1(\lambda_{0p} r)}y_1(\lambda_{0p} \rho)\right)\left(\frac{\rho^2}{\lambda_{0p}}\right)\bigg|_r^a\nonumber\\
    &=  \left(j_1(\lambda_{0p} a)-\frac{j_1(\lambda_{0p} r)}{y_1(\lambda_{0p} r)}y_1(\lambda_{0p} a)\right) \left(\frac{a^2}{\lambda_{0p}}\right)\nonumber.
\end{align}
We evaluate the limit:
\begin{align}
     \lim_{r\rightarrow a} \frac{1}{v^2}&\left(j_1(\lambda_{0p} a)-\frac{j_1(\lambda_{0p} r)}{y_1(\lambda_{0p} r)}y_1(\lambda_{0p} a)\right)^2 \nonumber \\
     &= \left(\frac{1}{4 \pi a^4 \lambda_{0p} }\right)^2\frac{1}{y_1^2\left(\lambda_{0p} a\right)}.
\end{align}
We show that $y_1^{-2}(x) = P_0^2(x)x^4$:
\begin{align}
    y_1^2(x)P_0^2(x)x^4 &= y_1^2(x) \left(j_0(x) - \frac{j_1(x)}{y_1(x)}y_0(x)\right)^2 x^4\nonumber\\
    &= (j_0(x)y_1(x) - j_1(x)y_0(x))^2 x^4\nonumber \\ 
    &= \left(\frac{-1}{x^2}\right)^2 x^4\nonumber \\
    &= 1\nonumber
\end{align}
and therefore in the small $a-r$ limit,
\begin{align}
    \bigg|\frac{1}{v}\int_{\text{sensor}}\hspace{-0.6cm} \psi_{\ell m p}(\Vec{x}) d^3x \bigg|^2 = \delta_{0\ell}\delta_{0m} P_0^2(\lambda_{0p}a).
\end{align}
We then write the expression for the fractional variance 
\begin{align}
    \frac{\delta n^2}{\langle n \rangle ^2} 
    &= \frac{2}{V\langle c \rangle  D T}\sum_{p=1}\frac{g_{00p}}{k^2_{00p}}P_0^2(\lambda_{0p}a) \\
    &= \frac{2}{V\langle c \rangle  D T}\sum_{p=1}g_{00p}P_0^2(\lambda_{0p}a) \frac{1}{\lambda_{0p}^2}  \\
    \label{eq:IS-sum}
    &= \frac{1}{\pi \langle c \rangle  DT R}\sum_{p=1}\frac{P_0^2(z_p a/R)}{P_0^2(z_p) - \frac{a^3}{R^3}P_0^2(z_p a/R)}\frac{1}{z_p^2}\nonumber\\
\end{align} 
where we have used the definition of $g_{00p}$, $k^2_{00p} = \lambda_{0p}^2$, and defined $z_p\equiv \lambda_{0p}R$. In the limit $a\ll R$, $P_0\rightarrow j_0$, and the sum reduces to 
\begin{align}
   \frac{\delta n^2 }{\langle n \rangle ^2} &\approx \frac{1}{\pi \langle c \rangle  DT R} \sum_{p=1}  \frac{j_0^2(z_p a / R)}{j_0^2(z_p)}\frac{1}{z_p^2}
\end{align}
which is the same result as for the PS (Eq. \ref{eq:PS-sum}). Therefore the PS and IS are equivalent in the small $a$ limit. In the limit $(R-a)\ll R$, the curvature is large compared to the distance between the two domain boundaries, and the problem becomes a quasi-1D problem. The 1D result for the sensing precision of a thin receptor placed next to one domain boundary, in a domain of length $L$ (Table \ref{tab:1d-precision}), is
\begin{align}
    \frac{\delta n^2 }{\langle n \rangle ^2} &= \frac{2L}{3 \langle c \rangle  D T}.
\end{align}
To map this result to the 3D problem, we replace $L$ by $R-a$ and the 1D concentration $\avg{c}$ by the concentration integrated over a sphere of radius $a$ in 3D, $4\pi a^2 \langle c \rangle $, so that
\begin{align}
\label{eq:IS-large}
    \frac{\delta n^2 }{\langle n \rangle ^2} &= \frac{2(R-a)}{3 (4\pi a^2 \langle c \rangle ) D T} = \frac{(R-a)}{6\pi a^2 \langle c \rangle  D T}.
\end{align}
We numerically evaluate the sum in Eq. \ref{eq:IS-sum}, and in Fig. \ref{fig:IS-interp}, demonstrate that it asymptotically approaches the limits expressed in equations \ref{eq:IS-large} and \ref{eq:PS-solution} in the large and small sensor limits, respectively.
\clearpage

\bibliography{bib}

\end{document}